\documentclass[12pt]{article}
\usepackage{amssymb,amsmath,epsfig}
\allowdisplaybreaks

\begin{document}

\title{\bf Energy Conditions in $f(\mathcal{G},T)$ Gravity}
\author{M. Sharif \thanks {msharif.math@pu.edu.pk} and Ayesha Ikram
\thanks{ayeshamaths91@gmail.com}\\
Department of Mathematics, University of the Punjab,\\
Quaid-e-Azam Campus, Lahore-54590, Pakistan.}

\date{}
\maketitle

\begin{abstract}
The aim of this paper is to introduce a new modified gravity theory
named as $f(\mathcal{G},T)$ gravity ($\mathcal{G}$ and $T$ are the
Gauss-Bonnet invariant and trace of the energy-momentum tensor,
respectively) and investigate energy conditions for two
reconstructed models in the context of FRW universe. We formulate
general field equations, divergence of energy-momentum tensor,
equation of motion for test particles as well as corresponding
energy conditions. The massive test particles follow non-geodesic
lines of geometry due to the presence of extra force. We express
energy conditions in terms of cosmological parameters like
deceleration, jerk and snap parameters. The reconstruction technique
is applied to this theory using de Sitter and power-law cosmological
solutions. We analyze energy bounds and obtain feasible constraints
on free parameters.
\end{abstract}

\section{Introduction}

Current cosmic accelerated expansion has been affirmed from a
diverse set of observational data coming from several astronomical
evidences including supernova type Ia, large scale structure, cosmic
microwave background radiation etc \cite{1}. This expanding paradigm
is considered as a consequence of mysterious force dubbed as dark
energy (DE) which possesses negatively large pressure. Modified
theories of gravity are considered as the favorite candidates to
unveil the enigmatic nature of this energy. These modified theories
are usually developed by including scalar invariants and their
corresponding generic functions in the Einstein-Hilbert action.

A remarkably interesting gravity theory is the modified Gauss-Bonnet
(GB) theory. A linear combination of the form
\begin{equation}\nonumber
\mathcal{G}=R_{\alpha\beta\xi\eta}
R^{\alpha\beta\xi\eta}-4R_{\alpha\beta}R^{\alpha\beta}+R^2,
\end{equation}
where $R_{\alpha\beta\xi\eta},~R_{\alpha\beta}$ and $R$ represent
the Riemann tensor, Ricci tensor and Ricci scalar, respectively is
called a Gauss-Bonnet invariant $(\mathcal{G})$. It is a second
order Lovelock scalar invariant and thus free from spin-2 ghosts
instabilities \cite{3}. Gauss-Bonnet combination is a
four-dimensional topological invariant which does not involve in the
field equations. However, it provides interesting results in same
dimensions when either coupled with scalar field or arbitrary
function $f(\mathcal{G})$ is added to Einstein-Hilbert action
\cite{4}. The latter approach is introduced by Nojiri and Odintsov
known as $f(\mathcal{G})$ theory of gravity \cite{4a}. Like other
modified theories, this theory is an alternative to study DE and is
consistent with solar system constraints \cite{5}. In this context,
there is a possibility to discuss a transition from decelerated to
accelerated as well as from non-phantom to phantom phases and also
to explain the unification of early and late times accelerated
expansion of the universe \cite{6}.

The fascinating problem of cosmic accelerated expansion has
successfully been discussed by taking into account modified theories
of gravity with curvature matter coupling. The motion of test
particles is studied in $f(R)$ and $f(\mathcal{G})$ gravity theories
non-minimally coupled with matter Lagrangian density
$(\mathcal{L}_{m})$. Consequently, the extra force experienced by
test particles is found to be orthogonal to their four velocities
and the motion becomes non-geodesic \cite{7}. It is found that for
certain choices of $\mathcal{L}_{m}$, the presence of extra force
vanishes in non-minimal $f(R)$ model while it remains preserved in
non-minimal $f(\mathcal{G})$ model. The geodesic deviation is weaker
in $f(\mathcal{G})$ gravity for small curvatures as compared to
non-minimal $f(R)$ gravity. Nojiri et al. \cite{8} studied the
non-minimally coupling of $f(R)$ and $f(\mathcal{G})$ theories with
$\mathcal{L}_{m}$ and found that such coupling naturally unifies the
inflationary era with current cosmic accelerated expansion.

In order to describe some realistic matter distribution, certain
conditions must be imposed on the energy-momentum tensor
($T_{\alpha\beta}$) known as energy conditions. These conditions
originate from Raychaudhuri equations with the requirement that not
only the gravity is attractive but also the energy density is
positive. The null (NEC), weak (WEC), dominant (DEC) and strong
(SEC) energy conditions are the four fundamental conditions. They
play a key role to study the theorems related to singularity and
black hole thermodynamics. Null energy condition is important to
discuss the second law of black hole thermodynamics while its
violation leads to Big-Rip singularity of the universe \cite{9}. The
proof of positive mass theorem is based on DEC \cite{10} while SEC
is useful to study Hawking-Penrose singularity theorem \cite{11}.

Energy conditions have been investigated in different modified
theories of gravity like $f(R)$ gravity, Brans-Dicke theory,
$f(\mathcal{G})$ gravity, generalized teleparallel theory \cite{12}.
Banijamali et al. \cite{13} investigated energy conditions for
non-minimally coupling $f(\mathcal{G})$ theory with
$\mathcal{L}_{m}$ and found that WEC is satisfied for specific
viable $f(\mathcal{G})$ models. Sharif and Waheed \cite{14} explored
energy bounds in the context of generalized second order
scalar-tensor gravity with the help of power-law ansatz for scalar
field. Sharif and Zubair \cite{15} derived these conditions in
$f(R,T,R_{\alpha\beta}T^{\alpha\beta})$ theory of gravity for two
specific models and also examined Dolgov-Kowasaki instability for
particular models of $f(R,T)$ gravity.

In this paper, we introduce a new modified theory of gravity named
as $f(\mathcal{G},T)$ gravity in which gravitational Lagrangian is
obtained by adding a generic function $f(\mathcal{G},T)$ in the
Einstein-Hilbert action. We study energy conditions for the
reconstructed $f(\mathcal{G},T)$ models using isotropic homogeneous
universe model. The paper has the following format. In section
\textbf{2}, we formulate the field equations of this gravity and
discuss the equation of motion for test particles while general
expressions for energy conditions as well as in terms of
cosmological parameters are discussed in section \textbf{3}. The
reconstruction of models and their energy bounds are analyzed in
section \textbf{4}. In the last section, we summarize our results.

\section{Field Equations of $f(\mathcal{G},T)$ Gravity}

In this section, we formulate the field equations for
$f(\mathcal{G},T)$ gravity. For this purpose, we assume action of
the following form
\begin{equation}\label{1}
\mathcal{S}=\frac{1}{2\kappa^2}\int
d^{4}x\sqrt{-g}[R+f(\mathcal{G},T)]+\int
d^4{x}\sqrt{-g}\mathcal{L}_{m},
\end{equation}
where $g$ and $\kappa$ represent determinant of the metric tensor
$(g_{\alpha\beta})$ and coupling constant, respectively. The
energy-momentum tensor is defined as \cite{16}
\begin{equation}\label{2}
T_{\alpha\beta}=-\frac{2}{\sqrt{-g}}\frac{\delta(\sqrt{-g}\mathcal{L}_{m})}{\delta
g^{\alpha\beta}}.
\end{equation}
Assuming that the matter distribution depends on the components of
$g_{\alpha\beta}$ but has no dependence on its derivatives, we
obtain
\begin{equation}\label{3}
T_{\alpha\beta}=g_{\alpha\beta
}\mathcal{L}_{m}-2\frac{\partial\mathcal{L}_{m}}{\partial
g^{\alpha\beta }}.
\end{equation}
The variation in the action (\ref{1}) gives
\begin{eqnarray}\nonumber
0&=&\delta\mathcal{S}=\frac{1}{2\kappa^2}\int
d^{4}x[(R+f(\mathcal{G},T))\delta\sqrt{-g}+\sqrt{-g}(\delta
R+f_{\mathcal{G}}(\mathcal{G},T)\delta\mathcal{G}\\\label{4}&+&
f_{T}(\mathcal{G},T)\delta T)]+\int
d^{4}x\delta(\sqrt{-g}\mathcal{L}_{m}),
\end{eqnarray}
where $f_{\mathcal{G}}(\mathcal{G},T)=\frac{\partial
f(\mathcal{G},T)}{\partial\mathcal{G}}$ and
$f_{T}(\mathcal{G},T)=\frac{\partial f(\mathcal{G},T)}{\partial T}$.
The variation of
$\sqrt{-g},\\~R^{\xi}_{\alpha\beta\eta},~R_{\alpha\eta}$ and $R$
provide the following expressions
\begin{eqnarray}\nonumber
\delta\sqrt{-g}&=&-\frac{1}{2}\sqrt{-g}g_{\alpha\beta}\delta
g^{\alpha\beta },\\\nonumber\delta
R^{\xi}_{\alpha\beta\eta}&=&\nabla_{\beta}(\delta\Gamma^{\xi}_{\eta\alpha})
-\nabla_{\eta}(\delta\Gamma^{\xi}_{\beta\alpha}),\\\nonumber&=&(g_{\alpha\lambda}
\nabla_{[\eta}\nabla_{\beta
]}+g_{\lambda[\beta}\nabla_{\eta]}\nabla_{\alpha}) \delta
g^{\xi\lambda}+\nabla_{[\eta}\nabla^{\xi}\delta
g_{\beta]\alpha},\\\label{4a}\delta R_{\alpha\eta}&=&\delta
R^{\xi}_{\alpha\xi\eta},\quad\delta
R=(R_{\alpha\beta}+g_{\alpha\beta}\nabla^2-\nabla_{\alpha}\nabla_{\beta})
\delta g^{\alpha\beta},
\end{eqnarray}
where $\Gamma^{\xi}_{\alpha\beta}$ and $\nabla_{\alpha}$ represent
the Christoffel symbol and covariant derivative, respectively. The
variation of $\mathcal{G}$ and $T$ yield
\begin{eqnarray}\nonumber\delta\mathcal{G}&=&2R\delta
R-4\delta(R_{\alpha\beta}R^{\alpha\beta})+\delta(R_{\alpha
\beta\xi\eta}R^{\alpha\beta\xi\eta}),
\\\label{4b}\delta T&=&(T_{\alpha\beta}+\Theta_{\alpha\beta})\delta
g^{\alpha\beta},\quad\Theta_{\alpha\beta}=g^{\xi\eta}\frac{\delta
T_{\xi\eta}}{\delta g_{\alpha\beta}}.
\end{eqnarray}

Using these variational relations in Eq.(\ref{4}), we obtain the
field equations of $f(\mathcal{G},T)$ gravity after simplification
as follows
\begin{eqnarray}\nonumber
G_{\alpha
\beta}&=&\kappa^2T_{\alpha\beta}-(T_{\alpha\beta}+\Theta_{\alpha\beta})f_{T}
(\mathcal{G},T)+\frac{1}{2}g_{\alpha
\beta}f(\mathcal{G},T)-(2RR_{\alpha\beta}-4R_{\alpha}^{\xi}R_{\xi\beta}
\\\nonumber&-&4R_{\alpha\xi\beta\eta}R^{\xi\eta}+2R_{\alpha
}^{\xi\eta\delta}
R_{\beta\xi\eta\delta})f_{\mathcal{G}}(\mathcal{G},T)-(2Rg_{\alpha\beta}
\nabla^2-2R\nabla_{\alpha}\nabla_{\beta}\\\nonumber&-&4g_{\alpha\beta}R^{\xi\eta}
\nabla_{\xi}\nabla_{\eta}-4R_{\alpha
\beta}\nabla^2+4R_{\alpha}^{\xi}\nabla_{\beta}\nabla_{\xi}
+4R_{\beta}^{\xi}\nabla_{\alpha}\nabla_{\xi}\\\label{5}&+&4R_{\alpha\xi\beta\eta}
\nabla^{\xi}\nabla^{\eta})f_{\mathcal{G}}(\mathcal{G},T),
\end{eqnarray}
where $G_{\alpha\beta}=R_{\alpha\beta}-\frac{1}{2}g_{\alpha\beta}R$
and $\nabla^2=\Box=\nabla_{\alpha}\nabla^{\alpha}$ denote the
Einstein tensor and d'Alembert operator, respectively. It is worth
mentioning here that for $f(\mathcal{G},T)=f(\mathcal{G})$,
Eq.(\ref{5}) reduces to the field equations for $f(\mathcal{G})$
gravity while $\Lambda(T)$ gravity ($\Lambda$ is the cosmological
constant) is obtained in the absence of quadratic invariant
$\mathcal{G}$ \cite{4a,16a}. Furthermore, the Einstein field
equations are recovered when $f(\mathcal{G},T)=0$. The trace of
Eq.(\ref{5}) is given by
\begin{eqnarray}\nonumber
&&R+\kappa^2T-(T+\Theta)f_{T}(\mathcal{G},T)+2f(\mathcal{G},T)
+2\mathcal{G}f_{\mathcal{G}}(\mathcal{G},T)-2R\nabla^2f_{\mathcal{G}}
(\mathcal{G},T)\\\nonumber&+&4R^{\alpha\beta}\nabla_{\alpha}\nabla_{\beta}
f_{\mathcal{G}}(\mathcal{G},T)=0,
\end{eqnarray}
where $\Theta=\Theta^{\alpha}_{\alpha}$. In this theory, the
covariant divergence of Eq.(\ref{5}) is non-zero given by
\begin{eqnarray}\nonumber
\nabla^{\alpha}T_{\alpha\beta}&=&\frac{f_{T}(\mathcal{G},T)}
{\kappa^2-f_{T}(\mathcal{G},T)}\left[(T_{\alpha\beta}
+\Theta_{\alpha\beta})\nabla^{\alpha}(\ln{f_{T}(\mathcal{G},T)})
+\nabla^{\alpha}\Theta_{\alpha\beta}\right.\\\label{5a}&-&\left.
\frac{1}{2}g_{\alpha\beta}\nabla^{\alpha}T\right].
\end{eqnarray}
To obtain a useful expression for $\Theta_{\alpha\beta}$, we
differentiate Eq.(\ref{3}) with respect to metric tensor
\begin{equation}\label{6}
\frac{\delta T_{\alpha\beta}}{\delta g^{\xi\eta}}=\frac{\delta
g_{\alpha\beta}}{\delta g^{\xi\eta}}\mathcal{L}_{m}+g_{\alpha\beta}
\frac{\partial\mathcal{L}_{m}}{\partial
g^{\xi\eta}}-2\frac{\partial^2\mathcal{L}_{m}}{\partial
g^{\xi\eta}\partial g^{\alpha\beta}}.
\end{equation}
Using the relations
\begin{equation}\nonumber
\frac{\delta g_{\alpha\beta}}{\delta
g^{\xi\eta}}=-g_{\alpha\mu}g_{\beta\nu}\delta_{\xi\eta}^{\mu\nu},\quad
\delta_{\xi\eta}^{\mu\nu}=\frac{\delta g^{\mu\nu}}{\delta
g^{\xi\eta}},
\end{equation}
where $\delta_{\xi\eta}^{\mu\nu}$ is the generalized Kronecker
symbol and Eq.(\ref{6}) in (\ref{4b}), we obtain
\begin{equation}\label{7}
\Theta_{\alpha\beta}=-2T_{\alpha\beta}+g_{\alpha\beta}\mathcal{L}_{m}-2g^{\xi\eta}
\frac{\partial^{2}\mathcal{L}_{m}}{\partial g^{\alpha\beta}\partial
g^{\xi\eta}}.
\end{equation}
This shows that once the value of $\mathcal{L}_{m}$ is determined,
we can find the expression for tensor $\Theta_{\alpha\beta}$.

We consider matter distribution as the perfect fluid given by
\begin{equation}\label{8}
T_{\alpha\beta}=(\rho+P)V_{\alpha}V_{\beta}-Pg_{\alpha\beta},
\end{equation}
where $\rho,~P$ and $V_{\alpha}$ are the density, pressure and four
velocity of the fluid, respectively. The four velocity satisfies the
relation $V_{\alpha}V^{\alpha}=1$ and the corresponding Lagrangian
density can be taken as $\mathcal{L}_{m}=-P$ \cite{17}. Thus
Eq.(\ref{7}) yields
\begin{equation}\label{9}
\Theta_{\alpha\beta}=-2T_{\alpha\beta}-Pg_{\alpha\beta}.
\end{equation}
Equation (\ref{5}) can be written in identical form to Einstein
field equations as
\begin{equation}\label{9a}
G_{\alpha\beta}=\kappa^2
T_{\alpha\beta}^{(eff)}=\kappa^2(T_{\alpha\beta}+T_{\alpha\beta}^{\mathcal{G}T}),
\end{equation}
where $T_{\alpha\beta}^{\mathcal{G}T}$ is the $f(\mathcal{G},T)$
contribution. For the case of perfect fluid, the expression for
$T_{\alpha\beta}^{\mathcal{G}T}$ is given by
\begin{eqnarray}\nonumber
T_{\alpha\beta}^{\mathcal{G}T}&=&\frac{1}{\kappa^2}\left[(\rho+P)V_{\alpha}V_{\beta}f_{T}
(\mathcal{G},T)+\frac{1}{2}g_{\alpha\beta}f(\mathcal{G},T)-(2RR_{\alpha\beta}-4R_{\alpha}^{\xi}R_{\xi\beta}
\right.\\\nonumber&-&\left.4R_{\alpha\xi\beta\eta}R^{\xi\eta}+2R_{\alpha}^{\xi\eta\delta}
R_{\beta\xi\eta\delta})f_{\mathcal{G}}(\mathcal{G},T)-(2Rg_{\alpha\beta}\nabla^2-2R\nabla_{\alpha}
\nabla_{\beta}\right.\\\nonumber&-&\left.4g_{\alpha\beta}R^{\xi\eta}\nabla_{\xi}\nabla_{\eta}
-4R_{\alpha\beta}\nabla^2+4R_{\alpha}^{\xi}\nabla_{\beta}\nabla_{\xi}
+4R_{\beta}^{\xi}\nabla_{\alpha}\nabla_{\xi}\right.
\\\label{9b}&+&\left.4R_{\alpha\xi\beta\eta}
\nabla^{\xi}\nabla^{\eta})f_{\mathcal{G}}(\mathcal{G},T)\right].
\end{eqnarray}
The line element for FRW universe model is
\begin{equation}\label{9c}
ds^2=dt^2-a^2(t)(dx^2+dy^2+dz^2),
\end{equation}
where $a(t)$ represents the scale factor. The corresponding field
equations are
\begin{equation}\label{9d}
3H^2=\kappa^2\rho_{eff},\quad -(2\dot{H}+3H^2)=\kappa^2P_{eff},
\end{equation}
where
\begin{eqnarray}\nonumber
\rho_{eff}&=&\rho+\frac{1}{\kappa^2}\left[(\rho+P)
f_{T}(\mathcal{G},T)+\frac{1}{2}f(\mathcal{G},T)-12H^2
(H^2+\dot{H})f_{\mathcal{G}}(\mathcal{G},T)\right.\\\label{9e}&+&\left.
12H^3\partial_{t}f_{\mathcal{G}}(\mathcal{G},T)\right],
\\\nonumber P_{eff}&=&P-\frac{1}{\kappa^2}
\left[\frac{1}{2}f(\mathcal{G},T)-12H^2(H^2+\dot{H})
f_{\mathcal{G}}(\mathcal{G},T)+8H(H^2+\dot{H})\right.
\\\label{9f}&\times&\left.\partial_{t}f_{\mathcal{G}}(\mathcal{G},T)
+4H^2\partial_{tt}f_{\mathcal{G}}(\mathcal{G},T)\right],
\end{eqnarray}
$\mathcal{G}=24H^2(H^2+\dot{H})$, $H=\dot{a}/a$ is the Hubble
parameter and dot represents the time derivative. The divergence of
$T_{\alpha\beta}$ takes the form
\begin{equation}\label{9g}
\dot{\rho}+3H(\rho+P)=\frac{-1}{\kappa^2+f_{T}(\mathcal{G},T)}\left[
\left(\dot{P}+\frac{1}{2}\dot{T}\right)f_{T}(\mathcal{G},T)+
(\rho+P)\partial_{t}f_{T}(\mathcal{G},T)\right].
\end{equation}
To obtain standard conservation equation
\begin{equation}\label{9h}
\dot{\rho}+3H(\rho+P)=0,
\end{equation}
we need an additional constraint by taking the right side of
Eq.(\ref{9g}) equal to zero given by
\begin{equation}\label{9k}
\left(\dot{P}+\frac{1}{2}\dot{T}\right)f_{T}(\mathcal{G},T)+
(\rho+P)\partial_{t}f_{T}(\mathcal{G},T)=0.
\end{equation}

Now, we briefly discuss the motion of test particles in
$f(\mathcal{G},T)$ gravity. For this purpose, using Eqs.(\ref{8})
and (\ref{9}) in (\ref{5a}), the divergence of energy-momentum
tensor for perfect fluid is given by
\begin{eqnarray}\nonumber
&&\nabla_{\beta}(\rho+P)V^{\alpha}V^{\beta}+(\rho+P)[V^{\beta}\nabla_{\beta}V^{\alpha}
+V^{\alpha}\nabla_{\beta}V^{\beta}]-g^{\alpha\beta}\nabla_{\beta}P
\\\nonumber&&=\frac{-2}{2\kappa^2+3f_{T}(\mathcal{G},T)}\left[T^{\alpha\beta}
\nabla_{\beta}f_{T}(\mathcal{G},T)+g^{\alpha\beta}\nabla_{\beta}(P
f_{T}(\mathcal{G},T))\right].
\end{eqnarray}
The contraction of above equation with projection operator
$(h_{\alpha\xi}=g_{\alpha\xi}-V_{\alpha\xi})$ gives the following
expression
\begin{equation}\label{a}
g_{\alpha\xi}V^{\beta}\nabla_{\beta}V^{\alpha}=\frac{(2\kappa^2
+f_{T}(\mathcal{G},T))\nabla_{\beta}}{(\rho+P)(2\kappa^2+3f_{T}
(\mathcal{G},T))}h_{\xi}^{\beta},
\end{equation}
where we have used the relations
$V^{\alpha}\nabla_{\beta}V_{\alpha}=0,~h_{\alpha\xi}V^{\alpha}=0$
and $h_{\alpha\xi}T^{\alpha\beta}=-Ph_{\xi}^{\beta}$. Multiplying
Eq.(\ref{a}) with $g^{\mu\xi}$ and using the following identity
\cite{17}
\begin{equation}\nonumber
V^{\beta}\nabla_{\beta}V^{\alpha}=\frac{d^2x^{\alpha}}{ds^2}+\Gamma^{\alpha}
_{\beta\xi}V^{\beta}V^{\xi},
\end{equation}
we obtain the equation of motion for massive test particles in this
gravity as
\begin{equation}\label{b}
\frac{d^2x^{\alpha}}{ds^2}+\Gamma^{\alpha}
_{\beta\xi}V^{\beta}V^{\xi}=\zeta^{\alpha},
\end{equation}
where
\begin{equation}\label{c}
\zeta^{\alpha}=\frac{(2\kappa^2+f_{T}(\mathcal{G},T))}{(\rho+P)(2\kappa^2+3f_{T}
(\mathcal{G},T))}(g^{\alpha\beta}-V^{\alpha}V^{\beta})\nabla_{\beta}P,
\end{equation}
represents the extra force acting on the test particles and is
perpendicular to the four velocity of the fluid
($\zeta^{\alpha}V_{\alpha}=0$). For pressureless fluid, Eq.(\ref{c})
gives $\zeta^{\alpha}=0$ and hence the dust particles follow the
geodesic trajectories both in general relativity as well as in
$f(\mathcal{G},T)$ gravity. The equation of motion for perfect fluid
in general relativity is recovered in the absence of coupling
between matter and geometry \cite{17a}.

\section{Energy Conditions}

Energy conditions are the coordinate invariant which incorporate the
common characteristics shared by almost every matter field. The
concept of energy conditions came from the Raychaudhuri equations
which play a key role to discuss the congruence of null and timelike
geodesics with the requirement that not only the gravity is
attractive but also the energy density is positive. These equations
describe the temporal evolution of expansion scalar $(\theta)$ as
follows \cite{18}
\begin{eqnarray}\label{10}
\frac{d\theta}{d\tau}&=&-\frac{1}{3}\theta^2+\omega_{\alpha\beta}
\omega^{\alpha\beta}-\sigma_{\alpha\beta}\sigma^{\alpha\beta}
-R_{\alpha\beta}u^{\alpha}u^{\beta},\\\label{11}\frac{d\theta}{d\tau}
&=&-\frac{1}{2}\theta^2+\omega_{\alpha\beta}\omega^{\alpha\beta}
-\sigma_{\alpha\beta}\sigma^{\alpha\beta}-R_{\alpha\beta}k^{\alpha}
k^{\beta},
\end{eqnarray}
where $\omega_{\alpha\beta},~\sigma_{\alpha\beta},~u^{\alpha}$ and
$k^{\alpha}$ represent the rotation, shear tensor, timelike and null
tangent vectors in the congruences, respectively. For non-geodesic
congruences, the temporal evolution of $\theta$ is affected by the
presence of acceleration term which arises due to non-gravitational
force like pressure gradient as \cite{18a}
\begin{equation}\label{11a}
\frac{d\theta}{d\tau}=-\frac{1}{3}\theta^2+\omega_{\alpha\beta}
\omega^{\alpha\beta}-\sigma_{\alpha\beta}\sigma^{\alpha\beta}
+\nabla_{\alpha}(V^{\beta}\nabla_{\beta}V^{\alpha})
-R_{\alpha\beta}V^{\alpha}V^{\beta}.
\end{equation}
Neglecting the quadratic terms due to rotation-free as well as small
distortions described by $\sigma_{\alpha\beta}$, Eqs.(\ref{10}) and
(\ref{11}) yield
\begin{equation}\nonumber \theta=-\tau
R_{\alpha\beta}u^{\alpha}u^{\beta},\quad \theta=-\tau
R_{\alpha\beta}k^{\alpha}k^{\beta}.
\end{equation}
Using the condition for gravity to be attractive, i.e., $\theta<0$,
we obtain $R_{\alpha\beta}u^{\alpha}u^{\beta}\geq0$ and
$R_{\alpha\beta}k^{\alpha}k^{\beta}\geq0$. The equivalent form of
these inequalities can be obtained by the inversion of the Einstein
field equations as
\begin{equation}\nonumber
\left(T_{\alpha\beta}-\frac{1}{2}g_{\alpha\beta}T\right)u^{\alpha}u^{\beta}\geq0,\quad
\left(T_{\alpha\beta}-\frac{1}{2}g_{\alpha\beta}T\right)k^{\alpha}k^{\beta}\geq0.
\end{equation}

For perfect fluid matter distribution, these inequalities provide
the energy constraints defined by
\begin{itemize}
\item NEC:$\quad\rho+P\geq0$,
\item WEC:$\quad\rho+P\geq0,\quad\rho\geq0,$
\item SEC:$\quad\rho+P\geq0,\quad\rho+3P\geq0,$
\item DEC:$\quad\rho\pm P\geq0,\quad\rho\geq0.$
\end{itemize}
These conditions show that the violation of NEC leads to the
violation of all other conditions. Due to purely geometric nature of
Raychaudhuri equations, the concept of energy bounds in modified
theories of gravity can be extended with the assumption that the
total cosmic matter distribution acts like a perfect fluid. The
energy conditions can be formulated by replacing $\rho$ and $P$ with
$\rho_{eff}$ and $P_{eff}$, respectively. The geodesic lines of
geometry are followed by dust particles in $f(\mathcal{G},T)$
gravity, therefore we consider pressureless fluid to discuss the
energy conditions. These conditions take the following form
\begin{eqnarray}\nonumber
&&\textbf{NEC:}\quad\rho_{eff}+P_{eff}=\rho+\frac{1}{\kappa^2}
\left[\rho
f_{T}(\mathcal{G},T)+4H(H^2-2\dot{H})\partial_{t}f_{\mathcal{G}}
(\mathcal{G},T)\right.\\\label{12}&&\left.-4H^2\partial_{tt}
f_{\mathcal{G}}(\mathcal{G},T)\right]\geq0,
\\\nonumber&&\textbf{WEC:}\quad\rho_{eff}=\rho+\frac{1}{2\kappa^2}
\left[2\rho f_{T}(\mathcal{G},T)+f(\mathcal{G},T)
-24H^2(H^2+\dot{H})\right.\\\label{13}&&\left.\times
f_{\mathcal{G}}(\mathcal{G},T)+24H^3\partial_{t}f_{\mathcal{G}}
(\mathcal{G},T)\right]\geq0,\\\nonumber&&\textbf{SEC:}\quad
\rho_{eff}+3P_{eff}=\rho-\frac{1}{\kappa^2}\left[f(\mathcal{G},T)
-\rho f_{T}(\mathcal{G},T)-24H^2(H^2+\dot{H})
\right.\\\label{14}&&\left.\times
f_{\mathcal{G}}(\mathcal{G},T)+12H(H^2+2\dot{H})\partial_{t}
f_{\mathcal{G}}(\mathcal{G},T)+12H^2\partial_{tt}f_{\mathcal{G}}
(\mathcal{G},T)\right]\geq0,\\\nonumber&&\textbf{DEC:}\quad\rho_{eff}
-P_{eff}=\rho+\frac{1}{\kappa^2}\left[\rho f_{T}(\mathcal{G},T)
+f(\mathcal{G},T)-24H^2(H^2+\dot{H})\right.\\\label{15}&&\left.
\times f_{\mathcal{G}}(\mathcal{G},T)+4H(5H^2+2\dot{H})\partial_{t}
f_{\mathcal{G}}(\mathcal{G},T)+4H^2\partial_{tt}f_{\mathcal{G}}
(\mathcal{G},T)\right]\geq0.
\end{eqnarray}
The Hubble parameter, Ricci scalar, GB invariant and their
derivatives can be written in terms of cosmic parameters as
\begin{eqnarray}\nonumber
\dot{H}&=&-H^2(1+q),\quad \ddot{H}=H^3(j+3q+2),\\\label{16}\dddot{H}
&=&H^4(s-4j-3q^2-12q-6),\\\nonumber R&=&-6H^2(1-q),
\quad\dot{R}=-6H^3(j-q-2),\\\label{17}\ddot{R}&=&-6H^4(s+8q+q^2+6),
\\\nonumber\mathcal{G}&=&-24qH^4,\quad\dot{\mathcal{G}}=24H^5(j+3q+2q^2),
\\\label{18}\ddot{\mathcal{G}}&=&24H^6(s-6j-6qj-12q-15q^2-2q^3),
\end{eqnarray}
where $q,~j$ and $s$ denote the deceleration, jerk and snap
parameters, respectively and are defined as \cite{19}
\begin{equation}\label{18}
q=-\frac{1}{H^2}\frac{\ddot{a}}{a},\quad
j=\frac{1}{H^3}\frac{\dddot{a}}{a},\quad
s=\frac{1}{H^4}\frac{\ddddot{a}}{a}.
\end{equation}
The energy conditions (\ref{12})-(\ref{15}) in the form of above
parameters are
\begin{eqnarray}\nonumber
&&\textbf{NEC:}\quad\rho_{eff}+P_{eff}=\rho+\frac{1}{\kappa^2}\left[\rho
f_{T}+4H^3(3+2q)(f_{\mathcal{GG}}\dot{\mathcal{G}}+f_{\mathcal{G}T}\dot{T})
\right.\\\label{19}&&\left.-4H^2(f_{\mathcal{GGG}}\dot{\mathcal{G}}^2
+2f_{\mathcal{GG}T}\dot{\mathcal{G}}\dot{T}+f_{\mathcal{G}TT}\dot{T}^2
+f_{\mathcal{GG}}\ddot{\mathcal{G}}+f_{\mathcal{G}T}\ddot{T})\right]\geq0,
\\\nonumber&&\textbf{WEC:}\quad\rho_{eff}=\rho+\frac{1}{2\kappa^2}\left[f+2\rho
f_{T}+24qH^4f_{\mathcal{G}}+24H^3(f_{\mathcal{GG}}\dot{\mathcal{G}}
\right.\\\label{20}&&\left.+f_{\mathcal{G}T}\dot{T})\right]\geq0,
\\\nonumber&&\textbf{SEC:}\quad\rho_{eff}+3P_{eff}=\rho+
\frac{1}{\kappa^2}\left[-f+\rho f_{T}-24qH^4
f_{\mathcal{G}}+12H^3(1+2q)\right.\\\nonumber&&\left.\times
(f_{\mathcal{GG}}\dot{\mathcal{G}}+f_{\mathcal{G}T}
\dot{T})-12H^2(f_{\mathcal{GGG}}\dot{\mathcal{G}}^2+2f_{\mathcal{GG}T}
\dot{\mathcal{G}}\dot{T}+f_{\mathcal{G}TT}\dot{T}^2+f_{\mathcal{GG}}
\ddot{\mathcal{G}}\right.\\\label{21}&&\left.+f_{\mathcal{G}T}\ddot{T})
\right]\geq0,\\\nonumber&&\textbf{DEC:}\quad
\rho_{eff}-P_{eff}=\rho+\frac{1}{\kappa^2}\left[f+\rho f_{T}
+24qH^4f_{\mathcal{G}}+4H^3(3-2q)(f_{\mathcal{GG}}\dot{\mathcal{G}}
\right.\\\label{22}&&\left.+f_{\mathcal{G}T}\dot{T})
+4H^2(f_{\mathcal{GGG}}\dot{\mathcal{G}}^2+2f_{\mathcal{GG}T}
\dot{\mathcal{G}}\dot{T}+f_{\mathcal{G}TT}\dot{T}^2
+f_{\mathcal{GG}}\ddot{\mathcal{G}}+f_{\mathcal{G}T}\ddot{T})\right]\geq0.
\end{eqnarray}

\section{Reconstruction of $f(\mathcal{G},T)$ Models}

In this section, we use the reconstruction technique and discuss the
energy conditions for de Sitter and power-law universe models.

\subsection{de Sitter Universe Model}

This cosmological model explains the exponential expansion of the
universe with constant Hubble expansion rate. The scale factor is
defined as \cite{20}
\begin{equation}\label{23}
a(t)=a_{0}e^{H_{0}t},\quad H=H_{0},
\end{equation}
where $a_{0}$ is constant at $t_{0}$. The values of $R$ and GB
invariant are
\begin{equation}\label{24}
R=-12H_{0}^2,\quad\mathcal{G}=24H_{0}^4.
\end{equation}
For pressureless fluid, Eq.(\ref{9h}) gives the energy density of
the form
\begin{equation}\label{25}
\rho=\rho_{0}e^{-3H_{0}t}.
\end{equation}
The trace of energy-momentum tensor and its derivatives have the
following expressions
\begin{equation}\label{26}
T=\rho,\quad \dot{T}=-3H_{0}T,\quad\ddot{T}=9H_{0}^2T.
\end{equation}
Using Eqs.(\ref{23})-(\ref{26}) in Eq.(\ref{9d}), we obtain partial
differential equation
\begin{eqnarray}\nonumber
&&\kappa^2T+\frac{1}{2}f(\mathcal{G},T)-12H_{0}^4f_{\mathcal{G}}(\mathcal{G},T)
+Tf_{T}(\mathcal{G},T)-36
H_{0}^4T{f_{\mathcal{G}T}}(\mathcal{G},T)\\\label{27}&-&3H_{0}^2=0,
\end{eqnarray}
whose solution is given by
\begin{equation}\label{28}
f(\mathcal{G},T)=c_{1}c_{2}\left(e^{c_{1}\mathcal{G}}T^{\gamma_{1}}
+T^{\gamma_{2}}\right)+\gamma_{3}T+\gamma_{4},
\end{equation}
where $c_{i}$'s are integration constants and
\begin{eqnarray}\nonumber
\gamma_{1}=-\frac{1}{2}\left(\frac{1-24c_{1}H_{0}^4}{1-36c_{1}H_{0}^{4}}\right),
\quad\gamma_{2}=-\frac{1}{2},\quad
\gamma_{3}=-\frac{2}{3}\kappa^2,\quad\gamma_{4}=6H_{0}^2.
\end{eqnarray}
The additional constraint (\ref{9k}) becomes
\begin{eqnarray}\nonumber
c_{1}c_{2}\frac{(1-24c_{1}H_{0}^{4})(1-30c_{1}H_{0}^{4})}{(1-36c_{1}H_{0}^{4})^{2}}
e^{c_{1}\mathcal{G}}T^{\gamma_{1}}
+c_{1}c_{2}T^{\gamma_{2}}+\gamma_{3}T=0.
\end{eqnarray}
This equation splits Eq.(\ref{28}) into two $f(\mathcal{G},T)$
functions with some additional constant relations between the
coefficients. The reconstructed model (\ref{28}) can be written as a
combination of those functions. We analyze the energy conditions for
the $f(\mathcal{G},T)$ model given in Eq.(\ref{28}) instead of
analyzing them separately. Using model (\ref{28}) in energy
conditions (\ref{12})-(\ref{15}), it follows that
\begin{eqnarray}\nonumber
&&\textbf{NEC:}\quad\rho_{eff}+P_{eff}=\rho+\frac{1}{\kappa^2}
\left[\rho\left\{c_{1}c_{2}(\gamma_{1}e^{c_{1}\mathcal{G}}
T^{(\gamma_{1}-1)}+\gamma_{2}T^{(\gamma_{2}-1)})+\gamma_{3}\right\}
\right.\\\label{29}&&\left.+12c_{1}^2c_{2}\gamma_{1}H_{0}^{4}(1-3\gamma_{1})
e^{c_{1}\mathcal{G}}T^{\gamma_{1}}\right]\geq0,\\\nonumber
&&\textbf{WEC:}\quad\rho_{eff}=\rho+\frac{1}{2\kappa^2}
\left[2\rho\{c_{1}c_{2}(e^{c_{1}\mathcal{G}}\gamma_{1}T^{(\gamma_{1}-1)}
+\gamma_{2}T^{(\gamma_{2}-1)})+\gamma_{3}\}\right.\\\nonumber&&\left.
+\{c_{1}c_{2}(e^{c_{1}\mathcal{G}}T^{\gamma_{1}}+T^{\gamma_{2}})
+\gamma_{3}T+\gamma_{4}\}-24c_{1}^2c_{2}H_{0}^4e^{c_{1}\mathcal{G}}
T^{\gamma_{1}}(1+3\gamma_{1})\right]\geq0,\\\label{30}
\\\nonumber&&\textbf{SEC:}\quad\rho_{eff}+3P_{eff}=\rho
-\frac{1}{\kappa^2}\left[c_{1}c_{2}(e^{c_{1}\mathcal{G}}T^{\gamma_{1}}
+T^{\gamma_{2}})+\gamma_{3}T+\gamma_{4}-\rho\right.\\\nonumber&&\left.
\times\{c_{1}c_{2}\left(\gamma_{1}e^{c_{1}\mathcal{G}}
T^{(\gamma_{1}-1)}+\gamma_{2}T^{(\gamma_{2}-1)}\right)+\gamma_{3}\}-
12c_{1}^{2}c_{2}e^{c_{1}\mathcal{G}}H_{0}^{4}T^{\gamma_{1}}\right.
\\\label{31}&&\left.\times\left\{2+3\gamma_{1}-9\gamma_{1}^{2}
\right\}\right]\geq0,\\\nonumber&&\textbf{DEC:}\quad
\rho_{eff}-P_{eff}=\rho+\frac{1}{\kappa^2}[\rho
\{c_{1}c_{2}(e^{c_{1}\mathcal{G}}\gamma_{1}T^{(\gamma_{1}-1)}
+\gamma_{2}
T^{(\gamma_{2}-1)})+\gamma_{3}\}\\\nonumber&&+\{c_{1}c_{2}(e^{c_{1}\mathcal{G}}
T^{\gamma_{1}}+T^{\gamma_{2}})+\gamma_{3}T+\gamma_{4}\}-12c_{1}^2c_{2}
H_{0}^{4}e^{c_{1}\mathcal{G}}T^{\gamma_{1}}\\\label{32}&&\times
\{2+\gamma_{1} (5-3\gamma_{1})\}]\geq0.
\end{eqnarray}
\begin{figure}
\epsfig{file=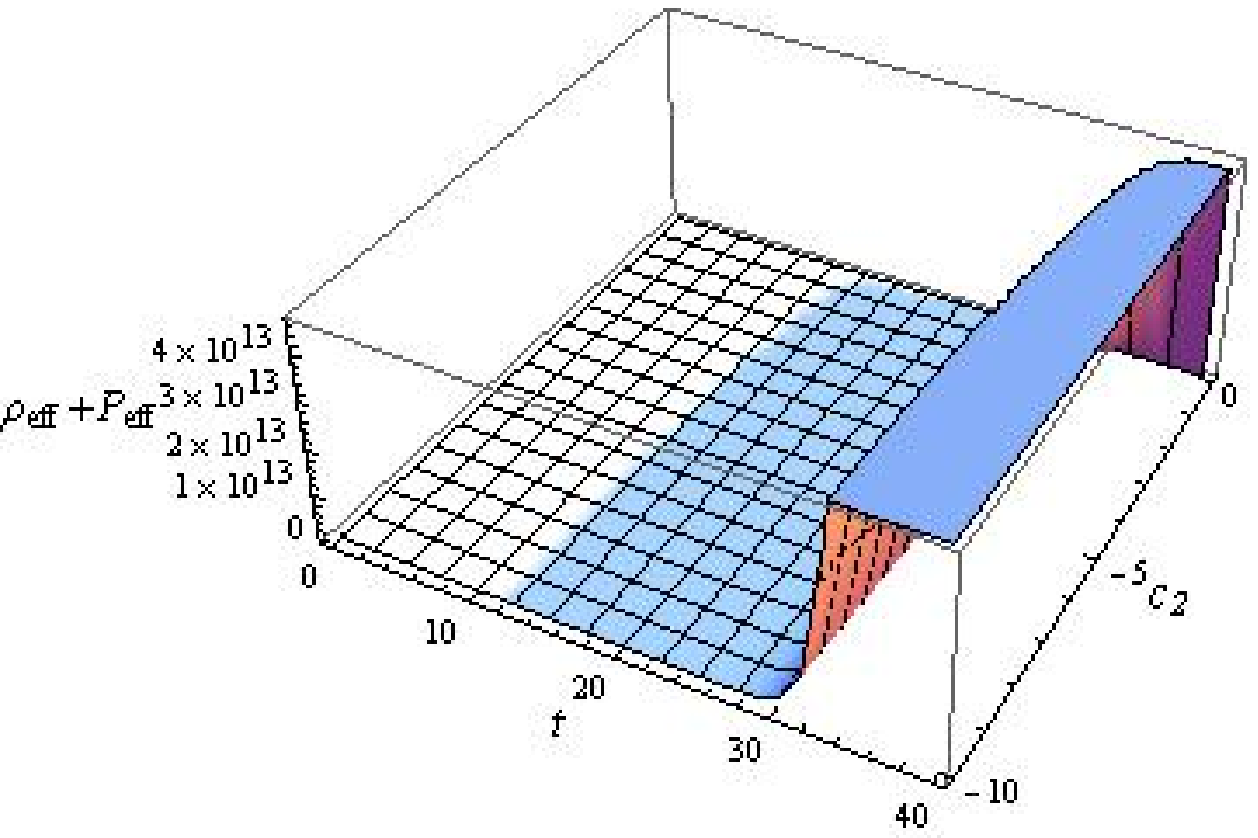, width=0.5\linewidth}\epsfig{file=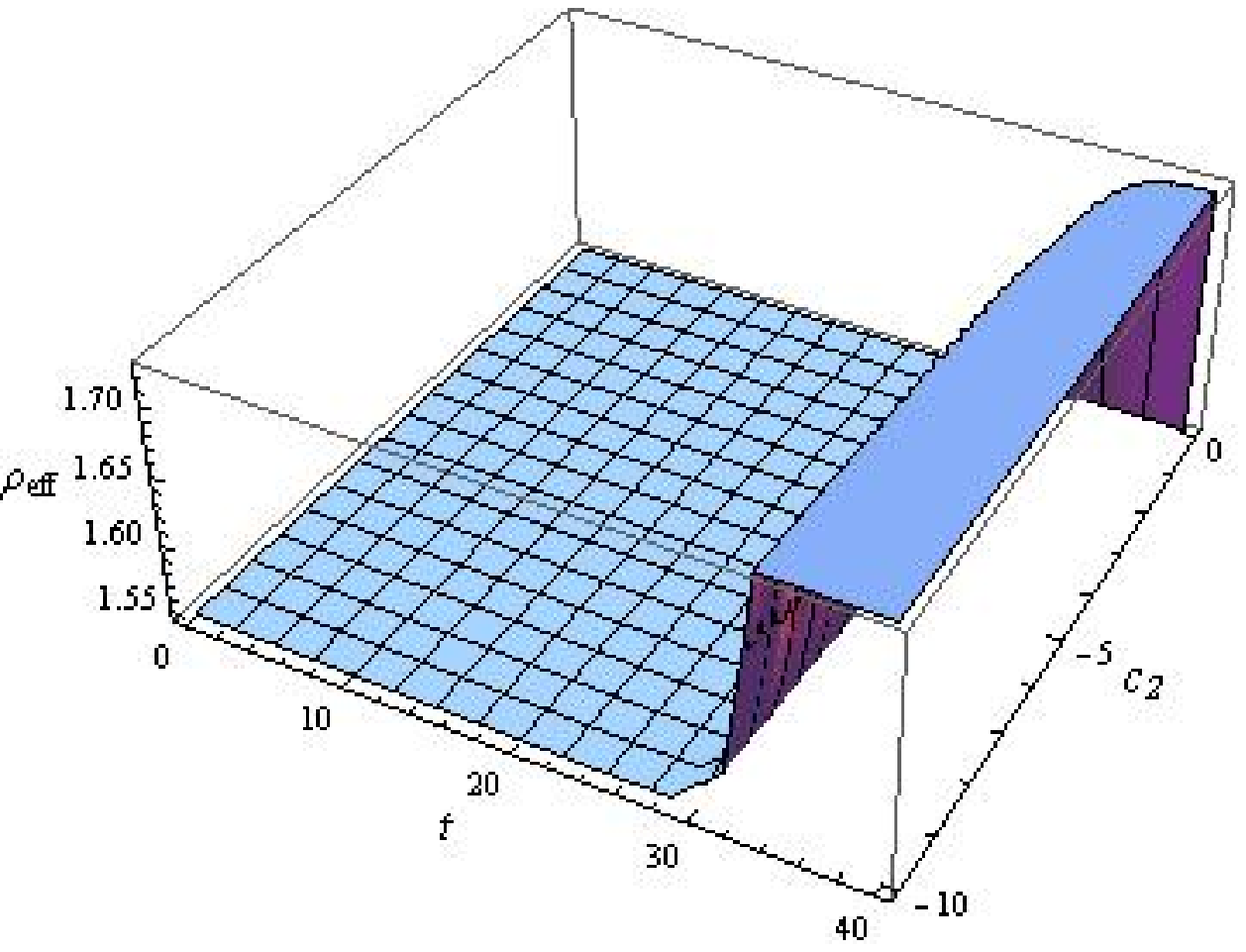,
width=0.5\linewidth}\caption{Energy conditions for $c_{1}=0.001$.}
\end{figure}
\begin{figure}
\epsfig{file=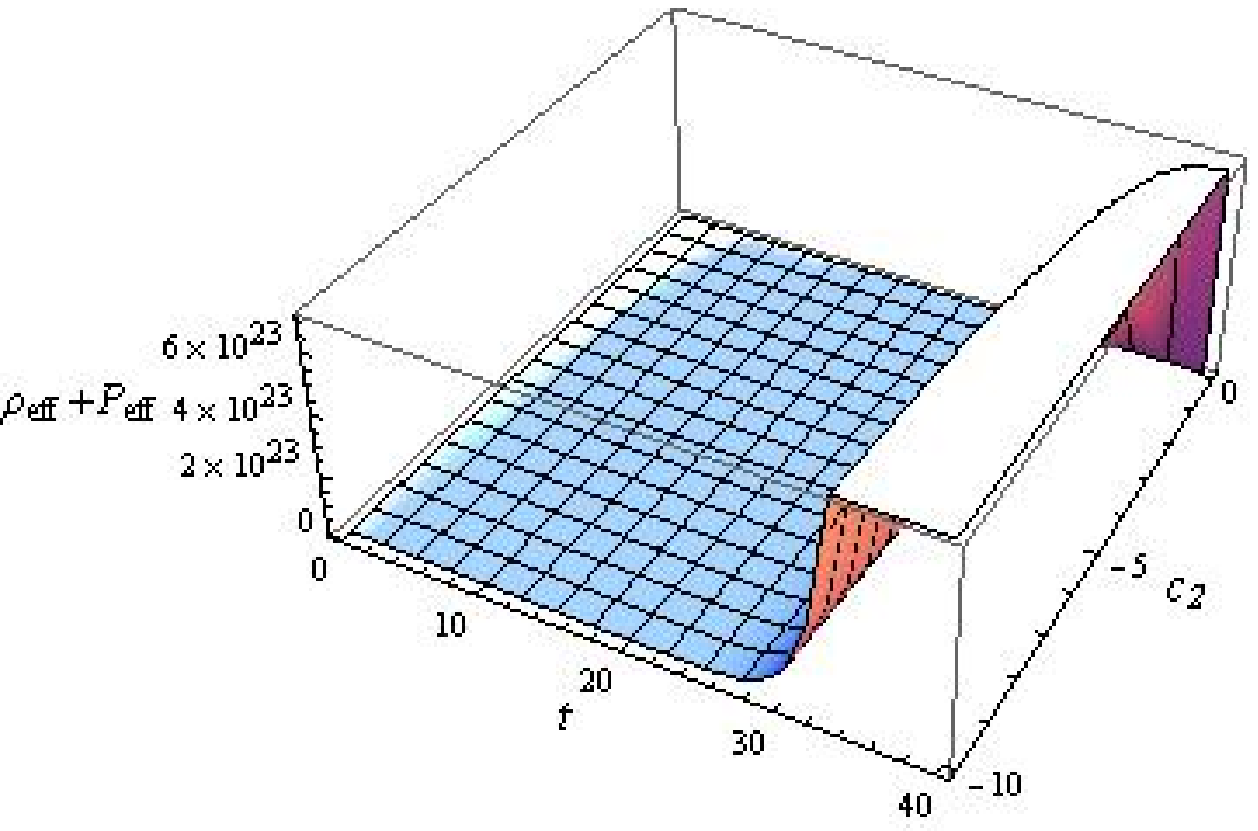, width=0.5\linewidth}\epsfig{file=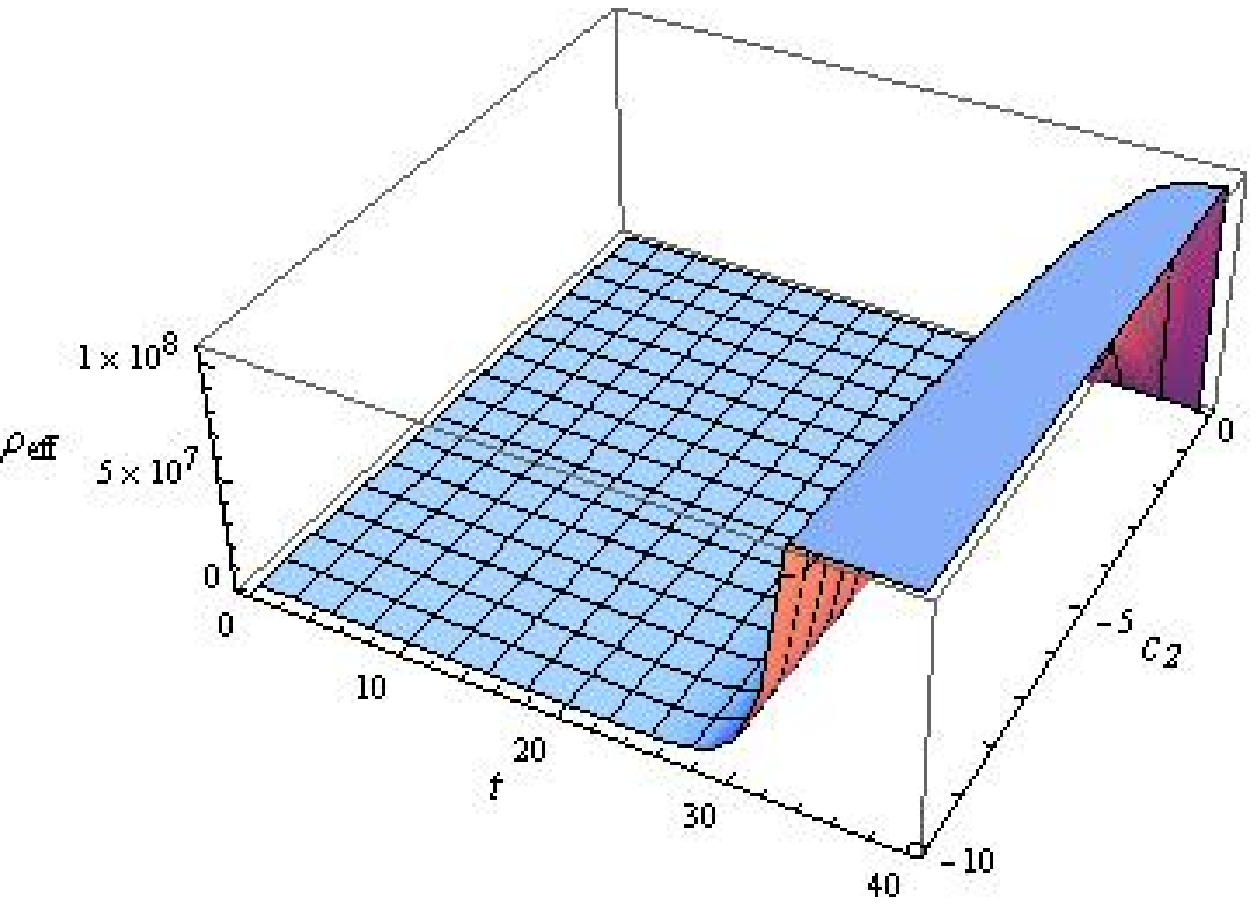,
width=0.5\linewidth}\caption{Energy conditions for $c_{1}=4$.}
\end{figure}

Figures \textbf{1} and \textbf{2} show the variation of NEC and WEC
for the case $c_{1}>0$ and $c_{2}<0$ with $\kappa=1$. We use the
following values of cosmological parameters:
$H_{0}=0.718,~q=-0.64,~j=1.02$ and $s=-0.39$ \cite{21}. In these
plots, we fix the constant $c_{1}$ for two arbitrarily chosen values
while $c_{2}$ varies from $[-10,0]$. Figure \textbf{1} shows the
positively increasing behavior of NEC as well as WEC with respect to
time in the considered interval of $c_{2}$. Figure \textbf{2} shows
similar behavior for $c_{1}=4$. In this case, both conditions are
satisfied for all values of $c_{1}$ and $c_{2}$. The energy
conditions for $(c_{1},c_{2})>0$ are discussed in Figures \textbf{3}
and \textbf{4}. The left plot of Figure \textbf{3} shows that the
NEC is satisfied for $t<3,~t<2.28$ and $t=2$ for $c_{2}=0.005,~0.05$
and $0.1$, respectively. Figure \textbf{4} (left) shows similar
decreasing behavior of time as the value of $c_{2}$ increases for
$c_{1}=0.01$. It is also observed that as the value of $c_{1}$
increases, the time interval for valid NEC decreases while the
positivity of $\rho_{eff}$ is shown in the right panel of both
figures. For the case $(c_{1},c_{2})>0$, both NEC and WEC are
satisfied for small values of $c_{1}$ and $c_{2}$ in a very small
time interval.
\begin{figure}
\epsfig{file=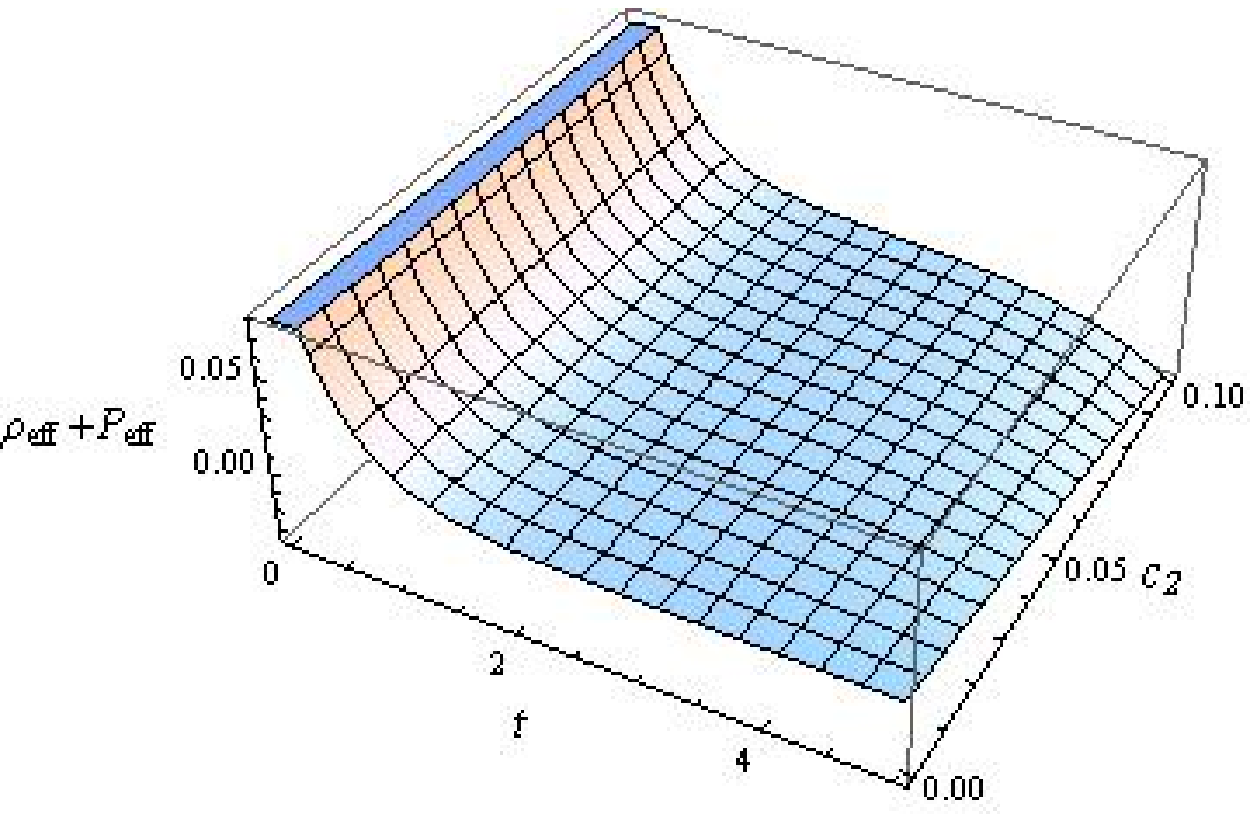, width=0.5\linewidth}\epsfig{file=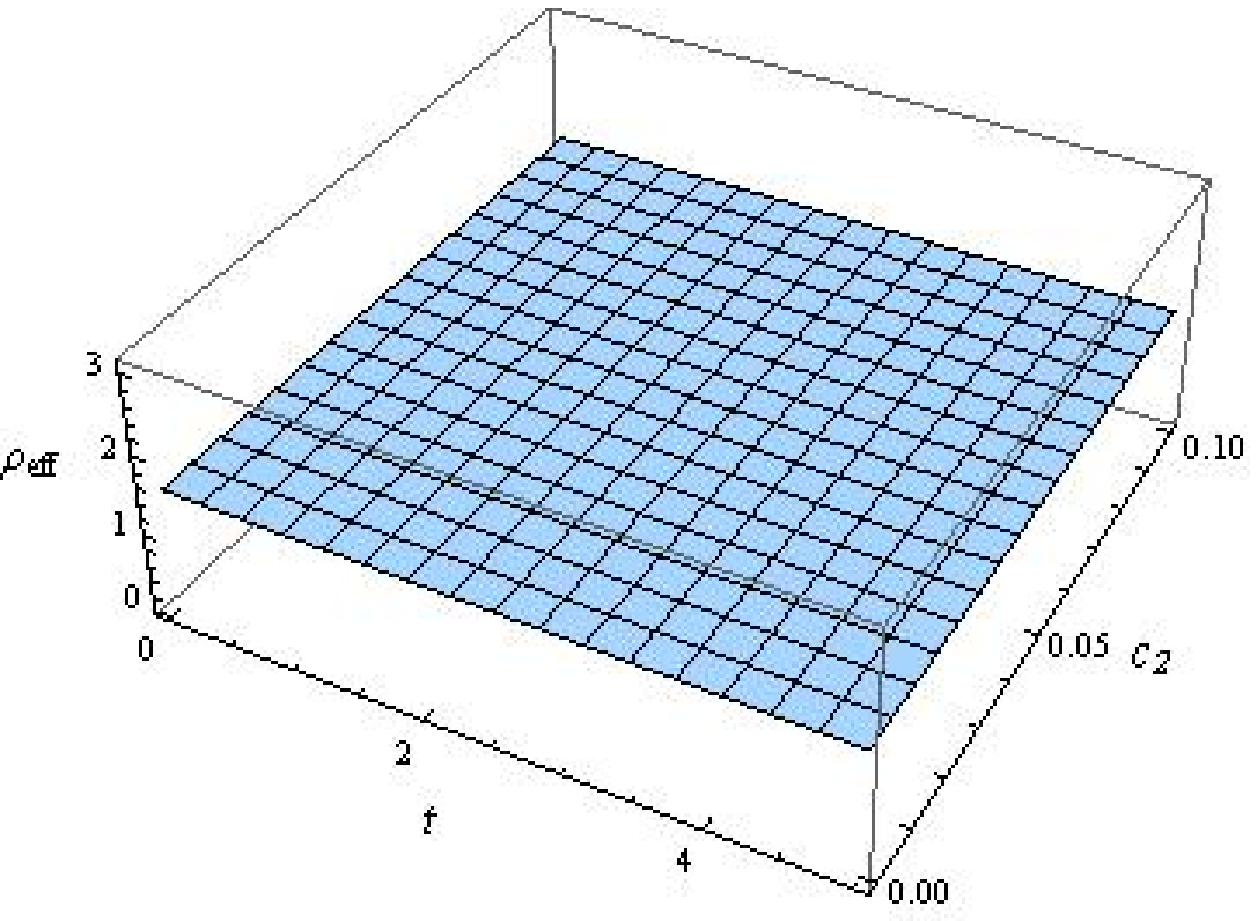,
width=0.5\linewidth}\caption{Energy conditions for $c_{1}=0.001$.}
\end{figure}
\begin{figure}
\epsfig{file=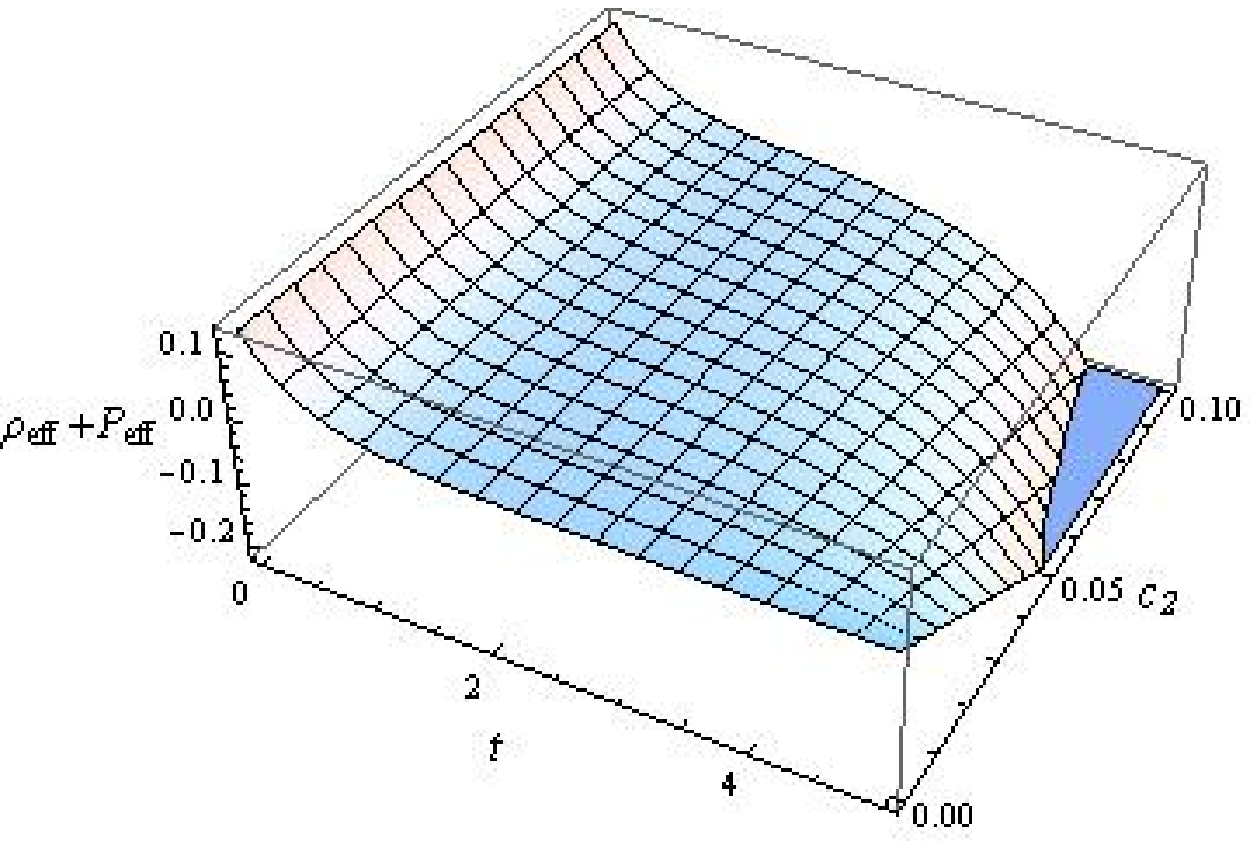, width=0.5\linewidth}\epsfig{file=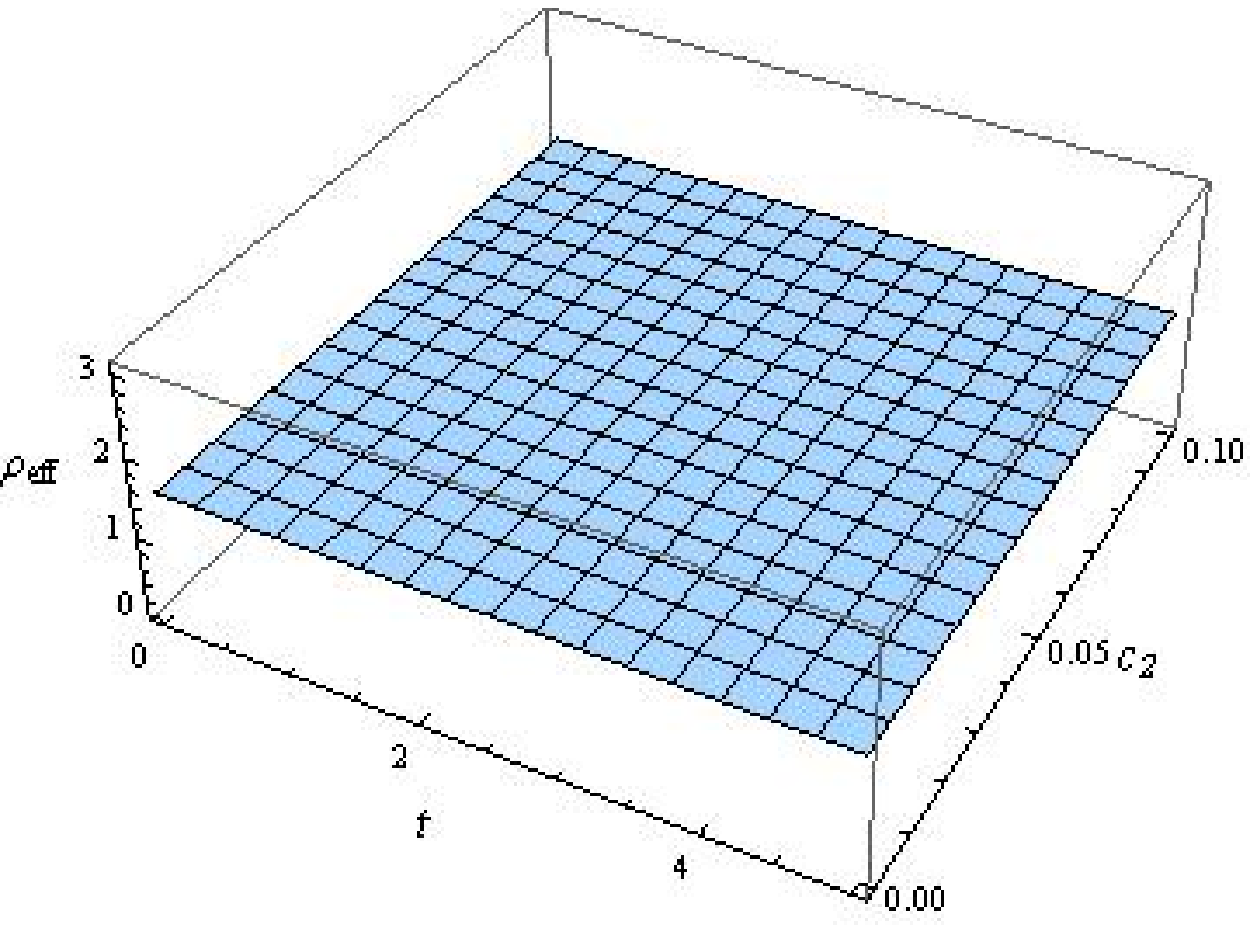,
width=0.5\linewidth}\caption{Energy conditions for $c_{1}=0.01$.}
\end{figure}
\begin{figure}
\epsfig{file=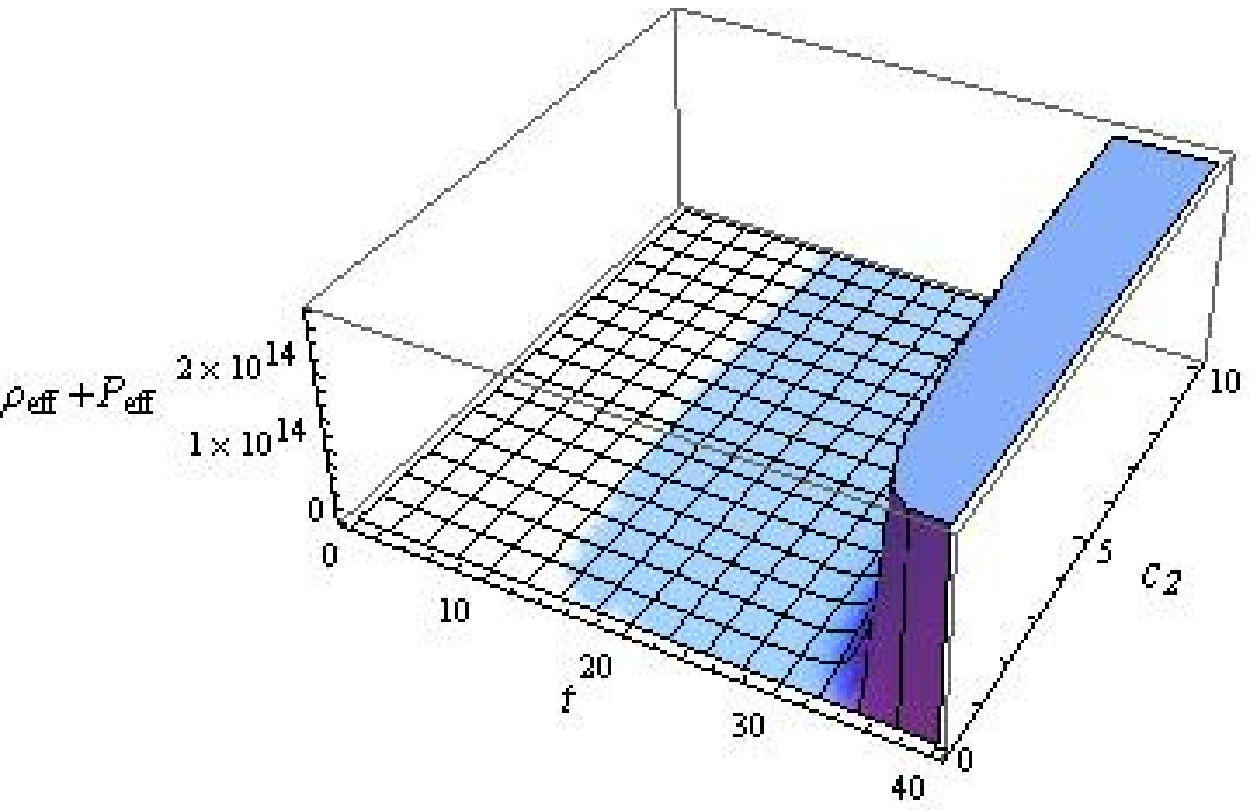, width=0.5\linewidth}\epsfig{file=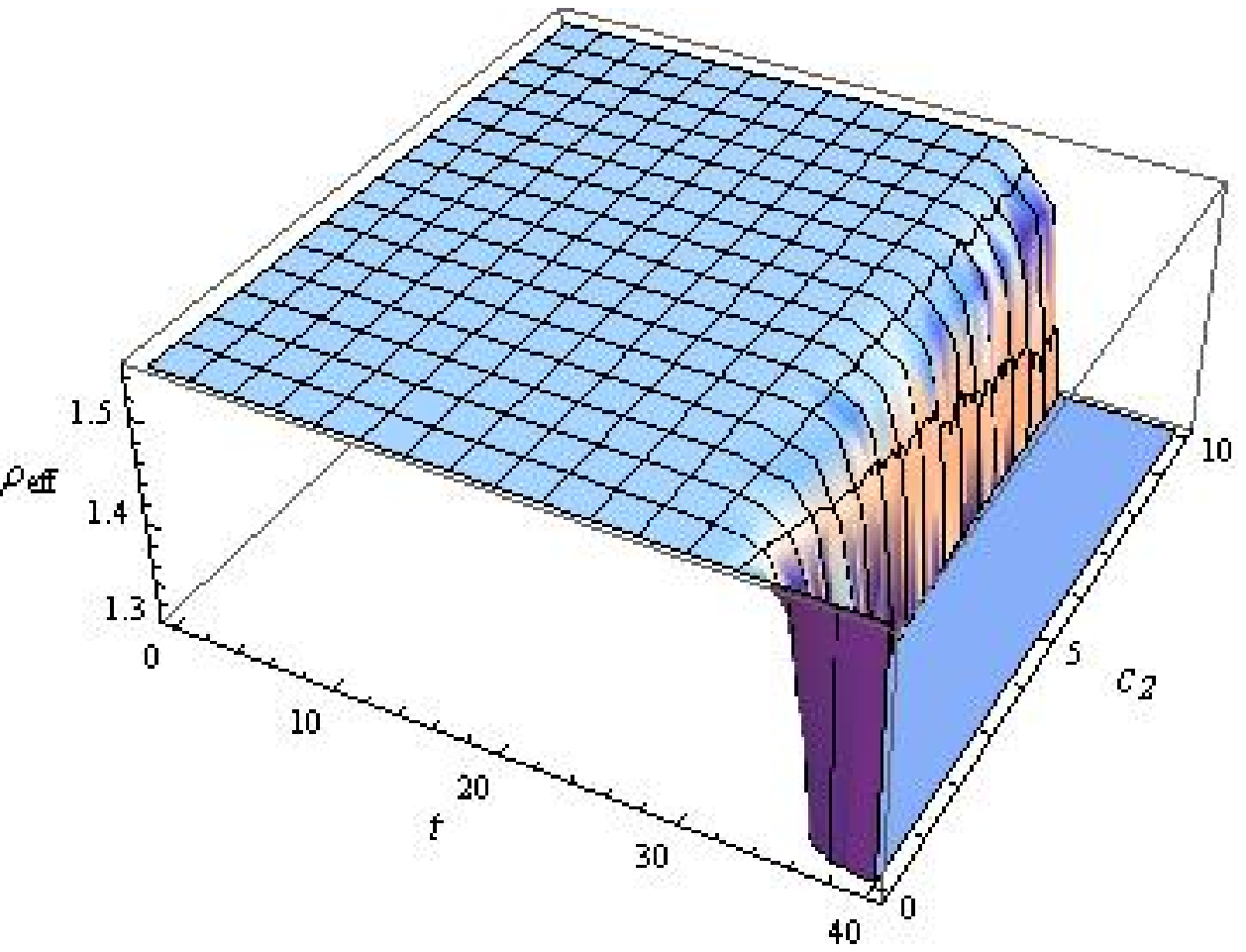,
width=0.5\linewidth}\caption{Energy conditions for $c_{1}=-0.01$.}
\end{figure}
\begin{figure}
\epsfig{file=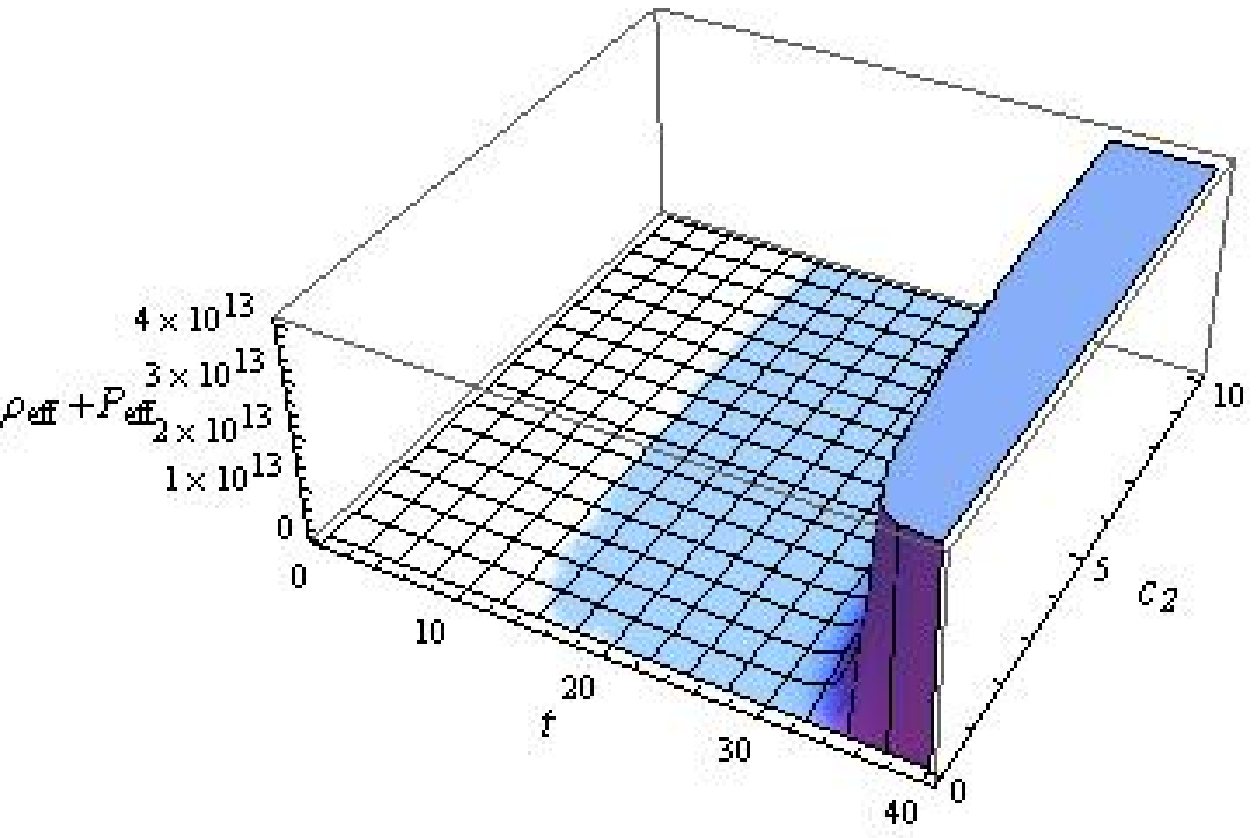, width=0.5\linewidth}\epsfig{file=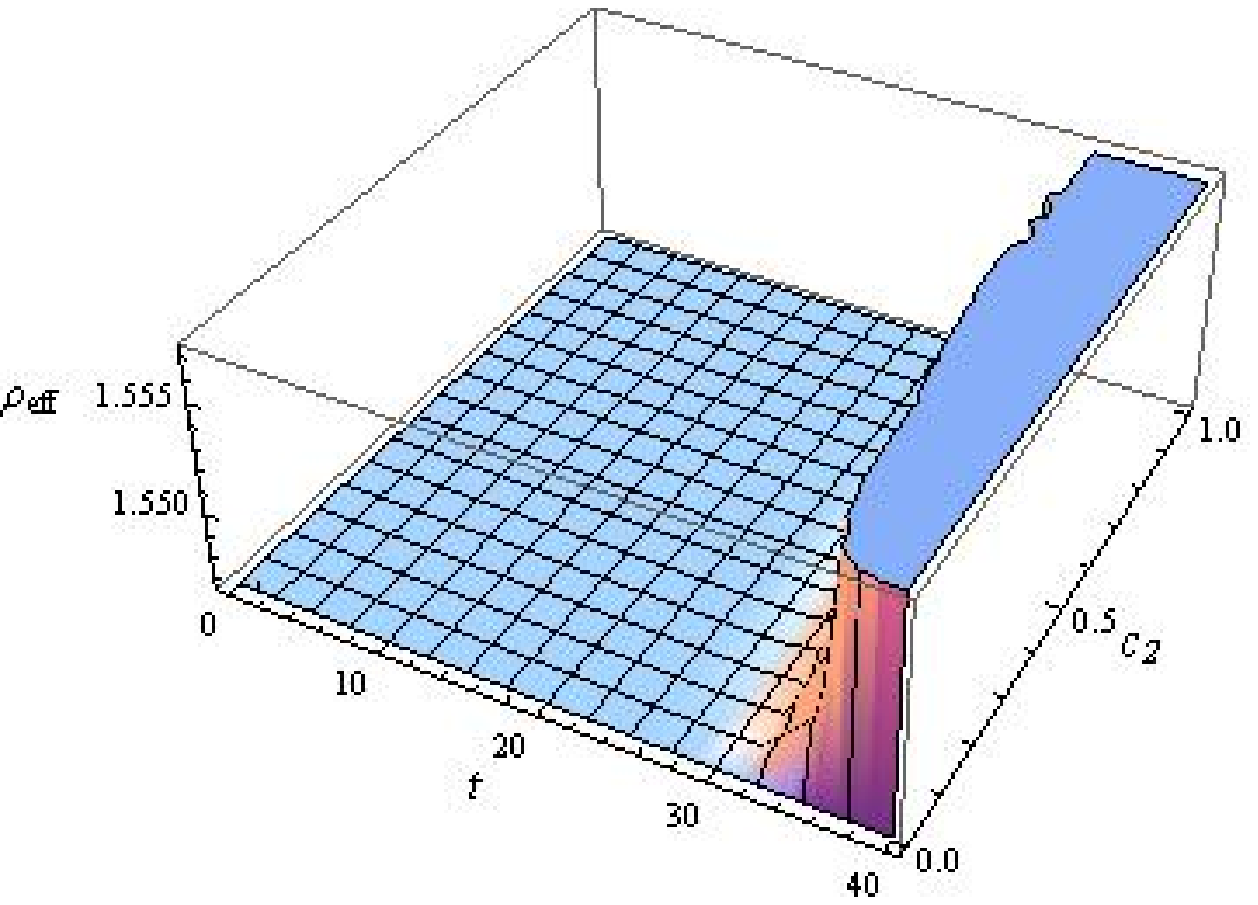,
width=0.5\linewidth}\caption{Energy conditions for $c_{1}=-0.001$.}
\end{figure}
\begin{figure}
\epsfig{file=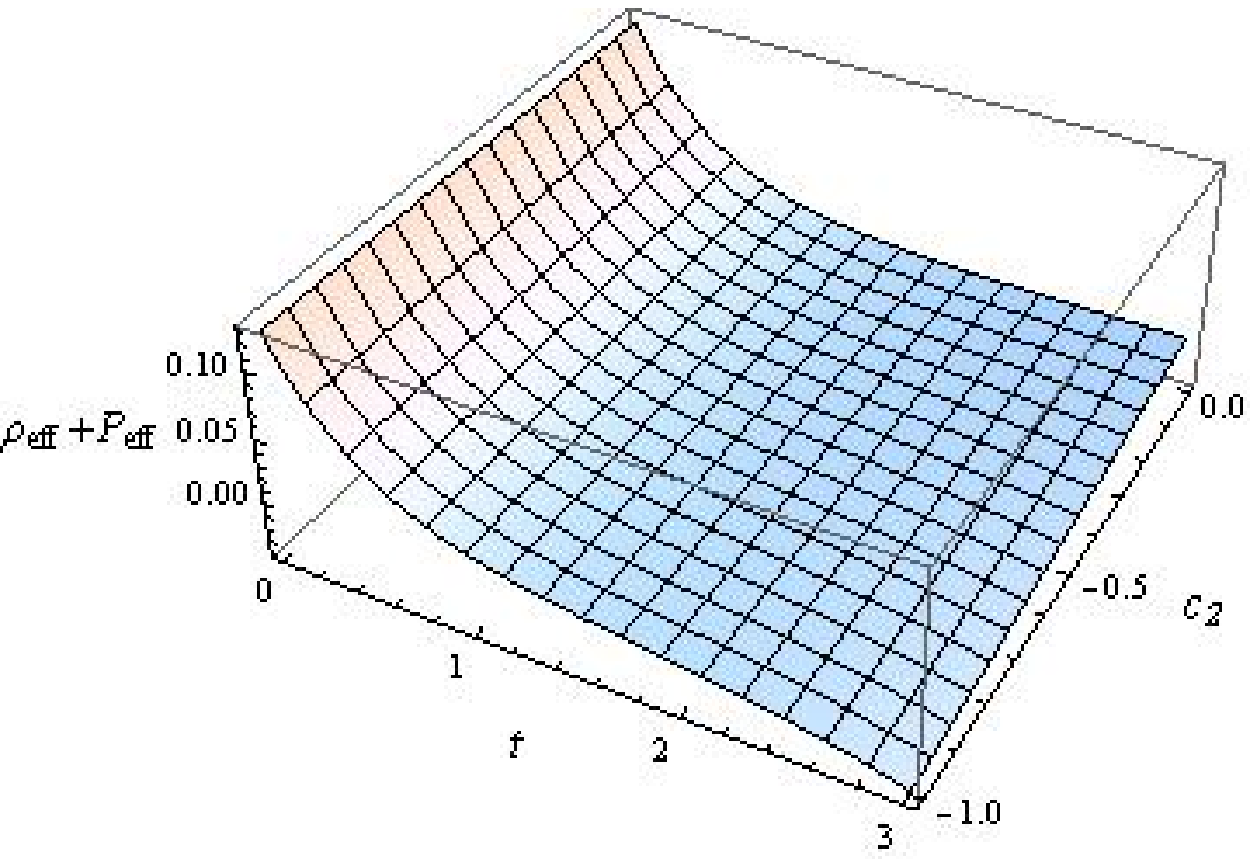, width=0.5\linewidth}\epsfig{file=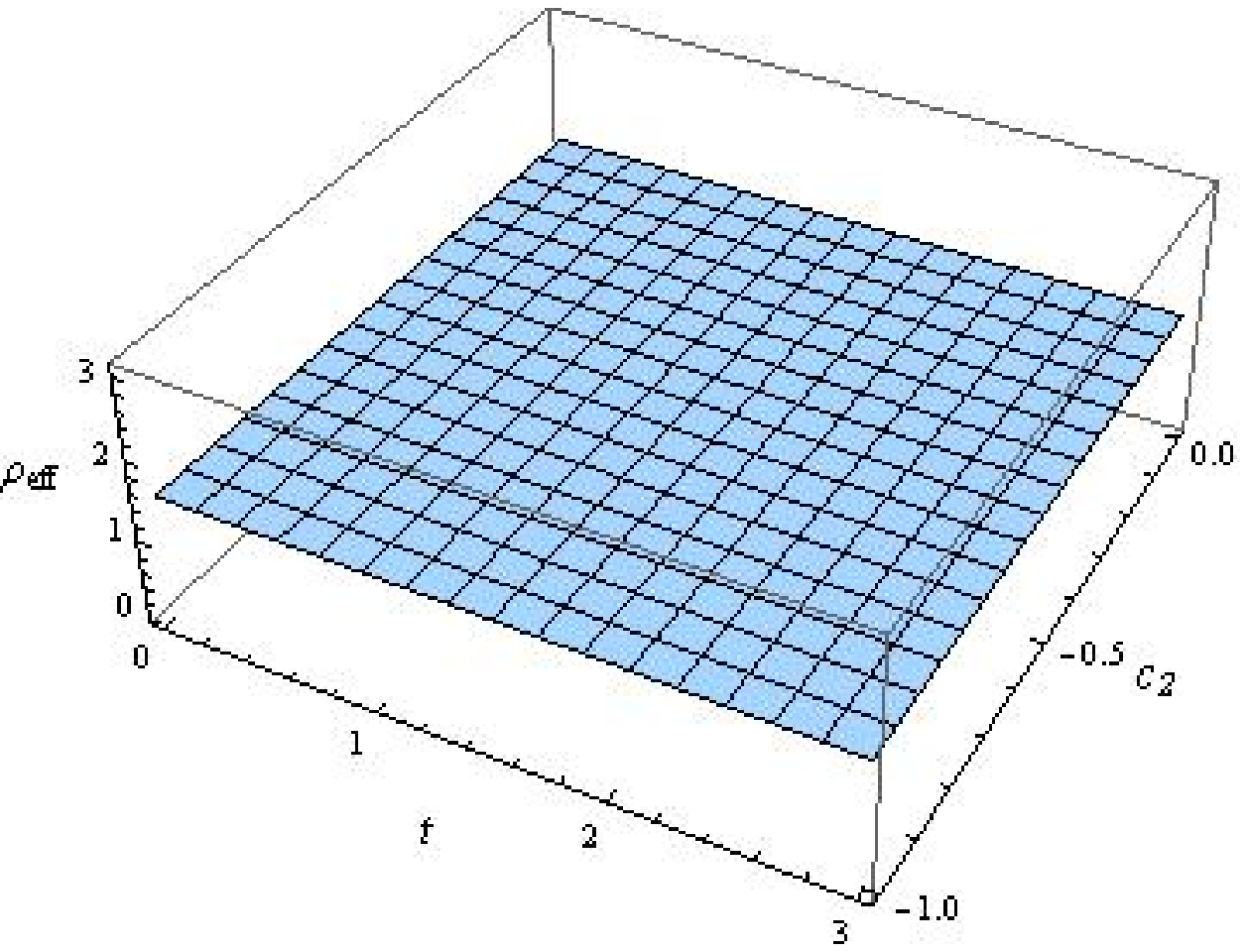,
width=0.5\linewidth}\caption{Energy conditions for $c_{1}=-0.001$.}
\end{figure}
\begin{figure}
\epsfig{file=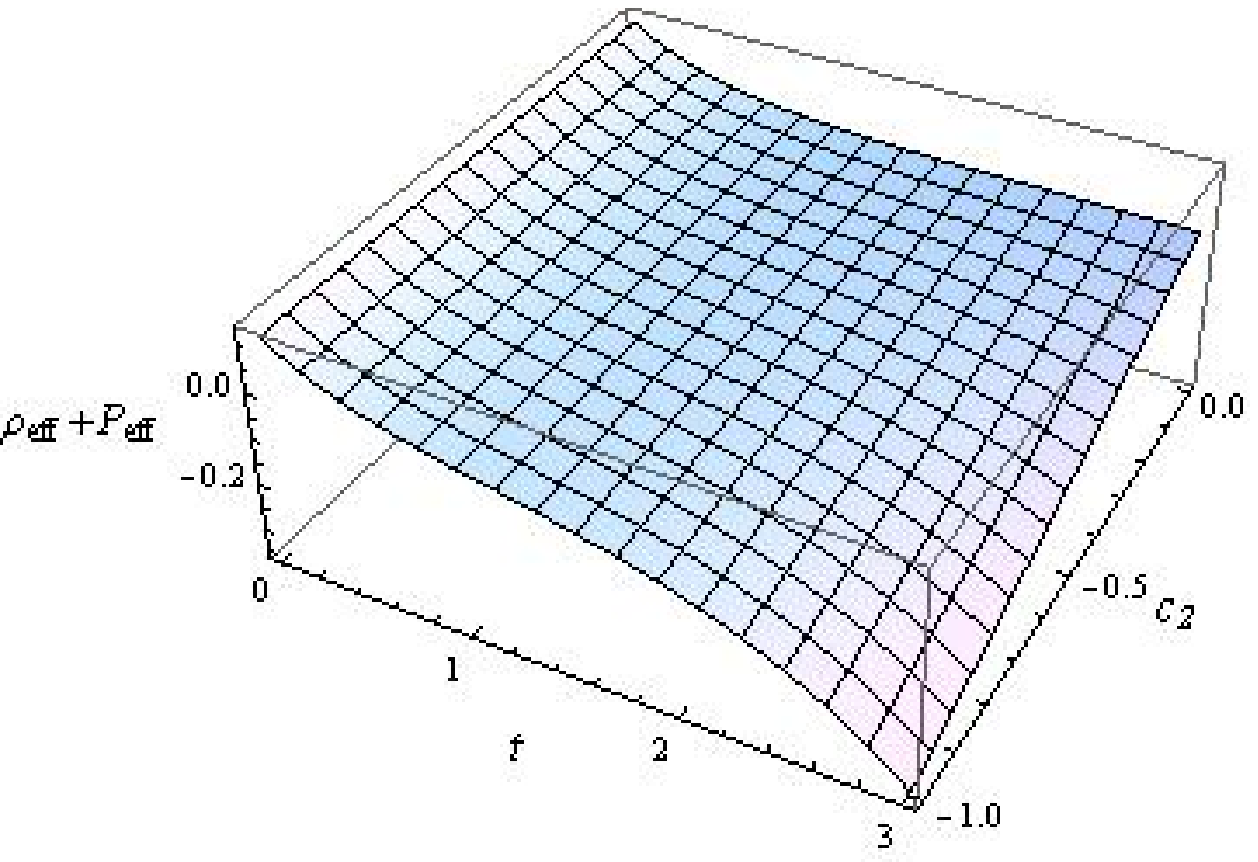, width=0.5\linewidth}\epsfig{file=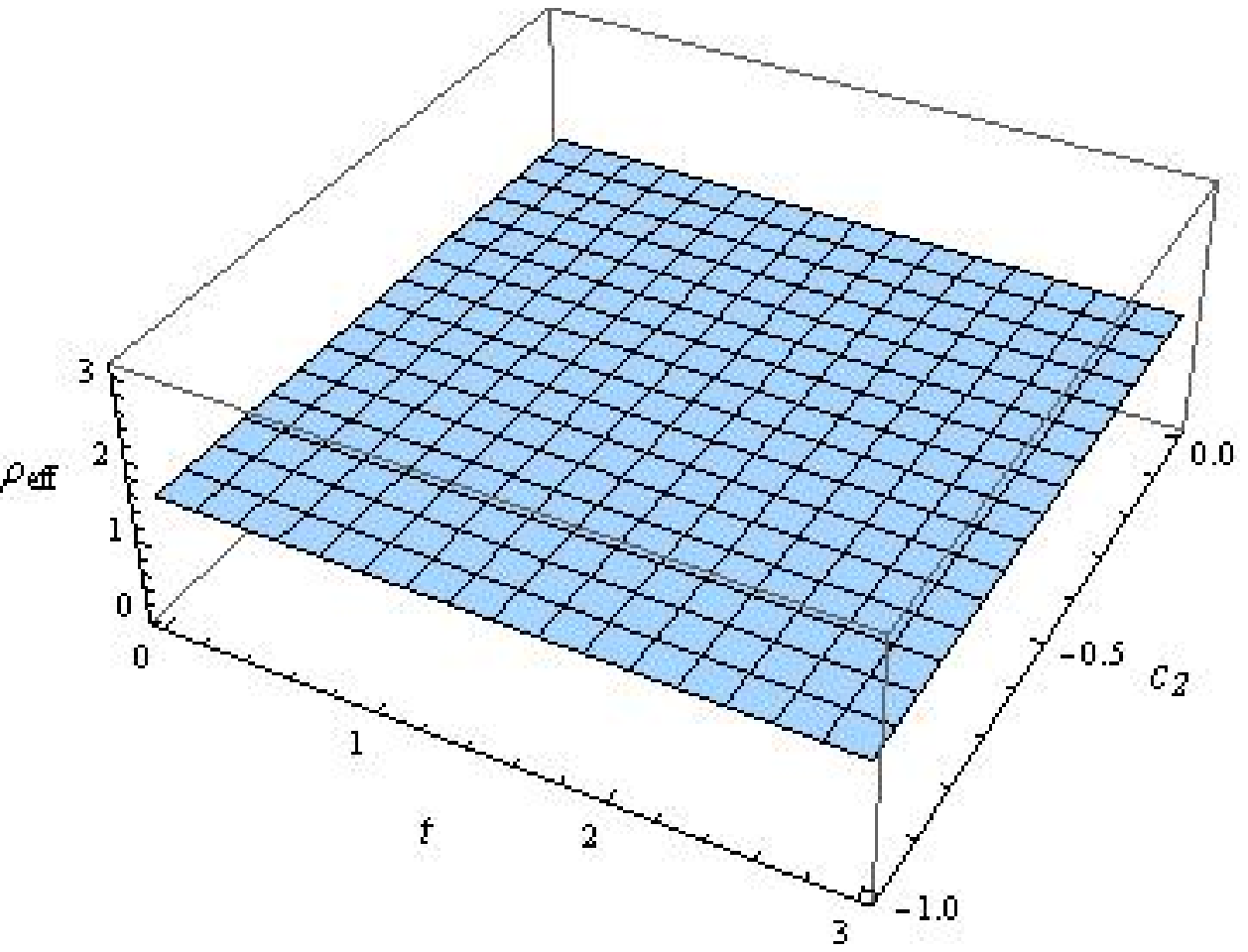,
width=0.5\linewidth}\caption{Energy conditions for $c_{1}=-0.01$.}
\end{figure}

Figures \textbf{5} and \textbf{6} deal with the case $c_{1}<0$ and
$c_{2}>0$. For arbitrarily chosen values of $c_{1}$, the increasing
behavior of NEC with respect to time is observed in the left panel
of both figures for all values of $c_{2}$. The right plot of Figure
\textbf{5} shows the positivity of $\rho_{eff}$ for $t<34$ while it
remains positive throughout the time interval for $c_{1}=-0.001$ as
shown in Figure \textbf{6} (right panel). The last possibility,
i.e., $c_{1}<0$ and $c_{2}<0$ is examined in Figures \textbf{7} and
\textbf{8}. The left panel of both figures show the decreasing and
increasing behavior of NEC as the time and integration constant
$c_{2}$ increase, respectively. The effective energy density
exhibits constant behavior for assumed values of $c_{1}$ in the
considered interval of $c_{2}$.

\subsection{Power-law Solution}

The power-law solution is of great interest to discuss the cosmic
evolution and its scale factor is defined as \cite{20}
\begin{equation}\label{33}
a(t)=a_{0}t^n,\quad H=\frac{n}{t},
\end{equation}
where $n>0$. For $0<n<1$, we have decelerated universe which leads
to radiation dominated era for $n=\frac{1}{2}$ and dust dominated
era for $n=\frac{2}{3}$ while cosmic accelerated era is observed for
$n>1$. The Ricci scalar and GB invariant are
\begin{equation}\label{34}
R=\frac{6n}{t^2}(1-2n),\quad\mathcal{G}=\frac{24n^3}{t^4}(n-1).
\end{equation}
The energy density for dust fluid is obtained from Eq.(\ref{9h}) as
\begin{equation}\label{35}
\rho=\rho_{0}t^{-3n}.
\end{equation}
The trace of $T_{\alpha\beta}$ and its time derivatives take the
form
\begin{equation}\label{35}
T=\rho,\quad\dot{T}=-\frac{3n}{t}T,
\quad\ddot{T}=\frac{3n}{t^2}(1+3n)T.
\end{equation}
Inserting Eqs.(\ref{33})-(\ref{35}) in the first field equation
(\ref{9d}), we obtain
\begin{eqnarray}\nonumber
&&\kappa^2T+\frac{1}{2}f(\mathcal{G},T)-\frac{1}{2}\mathcal{G}
f_{\mathcal{G}}(\mathcal{G},T)+Tf_{T}(\mathcal{G},T)-\left(
\frac{2}{n-1}\right)\mathcal{G}^2f_{\mathcal{GG}}(\mathcal{G},T)
\\\label{36}&-&\left(\frac{3n}{2(n-1)}\right)\mathcal{G}T
f_{\mathcal{G}T}(\mathcal{G},T)-3n^2\left(\frac{T}{\rho_{0}}
\right)^{\frac{2}{3n}}=0,
\end{eqnarray}
whose solution is given by
\begin{eqnarray}\nonumber
f(\mathcal{G},T)&=&d_{1}d_{3}T^{d_{2}}\mathcal{G}^{\frac{1}{4}
(\chi_{1}+\chi_{2})}+d_{2}d_{3}T^{d_{2}}\mathcal{G}^{-\frac{1}{4}
(\chi_{1}-\chi_{2})}+\chi_{3}T\\\label{37}&+&d_{1}d_{2}T^{\chi_{4}}
+\chi_{5}T^{\chi_{6}},
\end{eqnarray}
where $d_{i}$'s are constants of integration and
\begin{eqnarray}\nonumber
\chi_{1}&=&\frac{1}{2}\left[n^2(1+3d_{2}(3d_{2}+2))+2d_{2}(n-16)+3(2n+3)
\right]^{\frac{1}{2}},\\\nonumber\chi_{2}&=&\frac{1}{2}
\left[5-n(1+3d_{2})\right],\quad\chi_{3}=-\frac{2}{3}\kappa^2,\quad
\chi_{4}=-\frac{1}{2},
\\\nonumber\chi_{5}&=&\left(\frac{18n^3}{2+3n}\right)\rho_{0}
^{-\frac{2}{3n}},\quad\chi_{6} =\frac{2}{3n}.
\end{eqnarray}
In this case, Eq.(\ref{9k}) takes the form
\begin{eqnarray}\nonumber
&&d_{1}d_{3}T^{d_{2}}\mathcal{G}^{\frac{1}{4}
(\chi_{1}+\chi_{2})}\left[\frac{d_{2}}{6n}\{3n(2d_{2}-1)+2(\chi_{1}+\chi_{2})\}\right]
+d_{2}d_{3}T^{d_{2}}\mathcal{G}^{-\frac{1}{4}
(\chi_{1}-\chi_{2})}\\\nonumber&\times&\left[\frac{d_{2}}{6n}\{3n(2d_{2}-1)-2(\chi_{1}-\chi_{2})\}\right]
+\chi_{3}T+d_{1}d_{2}\chi_{4}^{2}T^{\chi_{4}}+\chi_{5}\chi_{6}^{2}T^{\chi_{6}}=0.
\end{eqnarray}
Solving Eq.(\ref{37}) with the above equation as in the previous
section, we obtain two functions whose combination is equivalent to
the reconstructed power-law $f(\mathcal{G},T)$ model.

Inserting the model (\ref{37}) in energy conditions
(\ref{19})-(\ref{22}), we obtain
\begin{eqnarray}\nonumber
&&\textbf{NEC:}\quad\rho_{eff}+P_{eff}=\rho+\frac{1}{\kappa^2}\left[
4H^3(3+2q)([\frac{1}{4}d_{1}d_{3}\right.\\\nonumber&&\left.\times
(\chi_{1}+\chi_{2})\left[\frac{1}{4}(\chi_{1}+\chi_{2})-1\right]
T^{d_{2}}\mathcal{G}^{\frac{1}{4}(\chi_{1}+\chi_{2})-2}
+\frac{1}{4}d_{2}d_{3}(\chi_{1}-\chi_{2})\right.\\\nonumber&&
\left.\times\left[\frac{1}{4}(\chi_{1}-\chi_{2})+1\right]
T^{d_{2}}\mathcal{G}^{-\frac{1}{4}(\chi_{1}-\chi_{2})-2}
]\dot{\mathcal{G}}
+[\frac{1}{4}d_{1}d_{2}d_{3}(\chi_{1}+\chi_{2})T^{d_{2}-1}\right.
\\\nonumber&&\left.\times\mathcal{G}^{\frac{1}{4}(\chi_{1}+\chi_{2})-1}
-\frac{1}{4}d_{2}^2d_{3}(\chi_{1}-\chi_{2})T^{d_{2}-1}\mathcal{G}^{-\frac{1}{4}
(\chi_{1}-\chi_{2})-1}]\dot{T})\right.\\\nonumber&&\left.-4H^2
([\frac{1}{4}d_{1}d_{3}(\chi_{1}+\chi_{2})\left[\frac{1}{4}(\chi_{1}+\chi_{2})-1\right]
\left[\frac{1}{4}(\chi_{1}+\chi_{2})-2\right]T^{d_{2}}\right.\\\nonumber&&
\left.\times\mathcal{G}^{\frac{1}{4}(\chi_{1}+\chi_{2})-3}-\frac{1}{4}d_{2}d_{3}
(\chi_{1}-\chi_{2})\left[\frac{1}{4}(\chi_{1}-\chi_{2})+1\right]\left[\frac{1}{4}
(\chi_{1}-\chi_{2})+2\right]\right.\\\nonumber&&\left.\times
T^{d_{2}}\mathcal{G}^{-\frac{1}{4}(\chi_{1}-\chi_{2})-3}]\dot{\mathcal{G}}^2
+2[\frac{1}{4}d_{1}d_{2}d_{3}(\chi_{1}+\chi_{2})\left[\frac{1}{4}(\chi_{1}
+\chi_{2})-1\right]T^{d_{2}-1}\right.\\\nonumber&&\left.\times\mathcal{G}^{\frac{1}{4}
(\chi_{1}+\chi_{2})-2}+\frac{1}{4}d_{2}^2d_{3}(\chi_{1}-\chi_{2})\left[\frac{1}{4}
(\chi_{1}-\chi_{2})+1\right]T^{d_{2}-1}\mathcal{G}^{-\frac{1}{4}(\chi_{1}-\chi_{2})-2}
]\right.\\\nonumber&&\left.\times\dot{\mathcal{G}}\dot{T}+[\frac{1}{4}d_{1}d_{2}d_{3}
(d_{2}-1)(\chi_{1}+\chi_{2})T^{d_{2}-2}\mathcal{G}^{\frac{1}{4}(\chi_{1}
+\chi_{2})-1}-\frac{1}{4}d_{2}^2d_{3}(d_{2}-1)\right.\\\nonumber&&\left.\times
(\chi_{1}-\chi_{2})T^{d_{2}-2}\mathcal{G}^{-\frac{1}{4}(\chi_{1}-\chi_{2})-1}
]\dot{T}^2+[\frac{1}{4}d_{1}d_{3}(\chi_{1}+\chi_{2})\left[\frac{1}{4}(\chi_{1}
+\chi_{2})-1\right]\right.\\\nonumber&&\left.\times
T^{d_{2}}\mathcal{G}^{\frac{1}{4}(\chi_{1}+\chi_{2})-2}
+\frac{1}{4}d_{2}d_{3}(\chi_{1}-\chi_{2})\left[\frac{1}{4}(\chi_{1}-\chi_{2})+1\right]
T^{d_{2}}\mathcal{G}^{-\frac{1}{4}(\chi_{1}-\chi_{2})-2}]\right.\\\nonumber&&\left.
\times\ddot{\mathcal{G}}+[\frac{1}{4}d_{1}d_{2}d_{3}(\chi_{1}+\chi_{2})T^{d_{2}-1}
\mathcal{G}^{\frac{1}{4}(\chi_{1}+\chi_{2})-1}-\frac{1}{4}
d_{2}^2d_{3}(\chi_{1}-\chi_{2})T^{d_{2}-1}\right.\\\nonumber&&\left.\times
\mathcal{G}^{-\frac{1}{4}(\chi_{1}-\chi_{2})-1}]\ddot{T})+\rho[d_{1}d_{2}
d_{3}T^{d_{2}-1}\mathcal{G}^{\frac{1}{4}(\chi_{1}+\chi_{2})}+d_{2}^2d_{3}
T^{d_{2}-1}\right.\\\label{38}&&\left.\times\mathcal{G}^{-\frac{1}{4}(\chi_{1}-\chi_{2})}
-\chi_{3}+d_{1}d_{2}\chi_{4}T^{\chi_{4}-1}+\chi_{5}\chi_{6}T^{\chi_{6}-1}
]\right]\geq0,\\\nonumber&&\textbf{WEC:}\quad\rho_{eff}=\rho+\frac{1}{2\kappa^2}
\left[d_{1}d_{3}T^{d_{2}}\mathcal{G}^{\frac{1}{4}(\chi_{1}+\chi_{2})}+
d_{2}d_{3}T^{d_{2}}\mathcal{G}^{-\frac{1}{4}(\chi_{1}-\chi_{2})}
\right.\\\nonumber&&\left.-\chi_{3}T+d_{1}d_{2}T^{\chi_{4}}+\chi_{5}T^{\chi_{6}}+2\rho
[d_{1}d_{2}d_{3}T^{d_{2}-1}\mathcal{G}^{\frac{1}{4}(\chi_{1}+\chi_{2})}+
d_{2}^2d_{3}T^{d_{2}-1}\right.\\\nonumber&&\left.\times\mathcal{G}^{-\frac{1}{4}
(\chi_{1}-\chi_{2})}-\chi_{3}+d_{1}d_{2}\chi_{4}T^{\chi_{4}-1}+\chi_{5}
\chi_{6}T^{\chi_{6}-1}]+24qH^4\right.\\\nonumber&&\left.\times
[\frac{1}{4}d_{1}d_{3}(\chi_{1}+\chi_{2})T^{d_{2}}\mathcal{G}^{\frac{1}{4}
(\chi_{1}+\chi_{2})-1}-\frac{1}{4}d_{2}d_{3}(\chi_{1}-\chi_{2})T^{d_{2}}
\mathcal{G}^{-\frac{1}{4}(\chi_{1}-\chi_{2})-1}]\right.\\\nonumber&&\left.
+24H^3([\frac{1}{4}d_{1}d_{3}(\chi_{1}+\chi_{2})\left[\frac{1}{4}(\chi_{1}
+\chi_{2})-1\right]T^{d_{2}}\mathcal{G}^{\frac{1}{4}(\chi_{1}+\chi_{2})-2}
+\frac{1}{4}d_{2}d_{3}\right.\\\nonumber&&\left.\times(\chi_{1}-\chi_{2})
\left[\frac{1}{4}(\chi_{1}-\chi_{2})+1\right]T^{d_{2}}\mathcal{G}^{-\frac{1}{4}
(\chi_{1}-\chi_{2})-2}]\dot{\mathcal{G}}+[\frac{1}{4}d_{1}d_{2}d_{3}(\chi_{1}+
\chi_{2})\right.\\\label{39}&&\left.\times T^{d_{2}-1}
\mathcal{G}^{\frac{1}{4}(\chi_{1}+\chi_{2})-1}-\frac{1}{4}
d_{2}^2d_{3}(\chi_{1}-\chi_{2})T^{d_{2}-1}\mathcal{G}^{-\frac{1}{4}(\chi_{1}-\chi_{2})-1}
]\dot{T})\right]\geq0,\\\nonumber
&&\textbf{SEC:}\quad\rho_{eff}+3P_{eff}=\rho+
\frac{1}{\kappa^2}\left[-[d_{1}d_{3}T^{d_{2}}\mathcal{G}
^{\frac{1}{4}(\chi_{1}+\chi_{2})}+d_{2}d_{3}T^{d_{2}}
\right.\\\nonumber&&\left.\times\mathcal{G}^{-\frac{1}{4}(\chi_{1}-\chi_{2})}
-\chi_{3}T+d_{1}d_{2}T^{\chi_{4}}+\chi_{5}T^{\chi_{6}}]
+\rho[d_{1}d_{2}d_{3}T^{d_{2}-1}\mathcal{G}^{\frac{1}{4}(\chi_{1}+\chi_{2})}
\right.\\\nonumber&&+\left.d_{2}^2d_{3}T^{d_{2}-1}\mathcal{G}
^{-\frac{1}{4}(\chi_{1}-\chi_{2})}-\chi_{3}+d_{1}d_{2}\chi_{4}
T^{\chi_{4}-1}+\chi_{5}\chi_{6}T^{\chi_{6}-1}]-24qH^4\right.
\\\nonumber&&\times\left.[\frac{1}{4}d_{1}d_{3}(\chi_{1}+\chi_{2})T^{d_{2}}
\mathcal{G}^{\frac{1}{4}(\chi_{1}+\chi_{2})-1}-\frac{1}{4}d_{2}d_{3}
(\chi_{1}-\chi_{2})T^{d_{2}}\mathcal{G}^{-\frac{1}{4}(\chi_{1}-\chi_{2})-1}]
\right.\\\nonumber&&+\left.12H^3(1+2q)([\frac{1}{4}d_{1}d_{3}
(\chi_{1}+\chi_{2})\left[\frac{1}{4}(\chi_{1}+\chi_{2})-1\right]
T^{d_{2}}\mathcal{G}^{\frac{1}{4}(\chi_{1}+\chi_{2})-2}
\right.\\\nonumber&&+\left.\frac{1}{4}d_{2}d_{3}(\chi_{1}-\chi_{2})
\left[\frac{1}{4}(\chi_{1}-\chi_{2})+1\right]T^{d_{2}}
\mathcal{G}^{-\frac{1}{4}(\chi_{1}-\chi_{2})-2}]\dot{\mathcal{G}}
+[\frac{1}{4}d_{1}d_{2}d_{3}\right.\\\nonumber&&\times\left.(\chi_{1}+\chi_{2})
T^{d_{2}-1}\mathcal{G}^{\frac{1}{4}(\chi_{1}+\chi_{2})-1}-\frac{1}{4}
d_{2}^2d_{3}(\chi_{1}-\chi_{2})T^{d_{2}-1}\mathcal{G}^{-\frac{1}{4}
(\chi_{1}-\chi_{2})-1}]\dot{T})\right.\\\nonumber&&-\left.12H^2([\frac{1}{4}
d_{1}d_{3}(\chi_{1}+\chi_{2})\left[\frac{1}{4}(\chi_{1}+\chi_{2})-1\right]
\left[\frac{1}{4}(\chi_{1}+\chi_{2})-2\right]T^{d_{2}}\right.\\\nonumber&&\left.
\times\mathcal{G}^{\frac{1}{4}(\chi_{1}+\chi_{2})-3}-\frac{1}{4}d_{2}d_{3}
(\chi_{1}-\chi_{2})\left[\frac{1}{4}(\chi_{1}-\chi_{2})+1\right]
\left[\frac{1}{4}(\chi_{1}-\chi_{2})+2\right]\right.\\\nonumber&&\left.\times
T^{d_{2}}\mathcal{G}^{-\frac{1}{4}(\chi_{1}-\chi_{2})-3}]\dot{\mathcal{G}}^2
+2[\frac{1}{4}d_{1}d_{2}d_{3}(\chi_{1}+\chi_{2})\left[\frac{1}{4}
(\chi_{1}+\chi_{2})-1\right]T^{d_{2}-1}\right.\\\nonumber&&\times\mathcal{G}^{\frac{1}{4}
(\chi_{1}+\chi_{2})-2}\left.+\frac{1}{4}d_{2}^2
d_{3}(\chi_{1}-\chi_{2})\left[\frac{1}{4}(\chi_{1}-\chi_{2})+1\right]
T^{d_{2}-1}\mathcal{G}^{-\frac{1}{4}(\chi_{1}-\chi_{2})-2}]\right.
\\\nonumber&&\left.\times\dot{\mathcal{G}}\dot{T}+[\frac{1}{4}d_{1}
d_{2}d_{3}(d_{2}-1)(\chi_{1}+\chi_{2})T^{d_{2}-2}
\mathcal{G}^{\frac{1}{4}(\chi_{1}+\chi_{2})-1}-\frac{1}{4}d_{2}^2
d_{3}(d_{2}-1)\right.\\\nonumber&&\left.\times(\chi_{1}-\chi_{2})T^{d_{2}-2}
\mathcal{G}^{-\frac{1}{4}(\chi_{1}-\chi_{2})-1}]\dot{T}^2+[\frac{1}{4}
d_{1}d_{3}(\chi_{1}+\chi_{2})\left[\frac{1}{4}(\chi_{1}+\chi_{2})
\right.\right.\\\nonumber&&\left.\left.-1\right]
T^{d_{2}}\mathcal{G}^{\frac{1}{4}(\chi_{1}+\chi_{2})-2}
+\frac{1}{4}d_{2}d_{3}(\chi_{1}-\chi_{2})\left[\frac{1}{4}
(\chi_{1}-\chi_{2})+1\right]T^{d_{2}}\right.\\\nonumber&&\left.\times
\mathcal{G}^{-\frac{1}{4}
(\chi_{1}-\chi_{2})-2}]\ddot{\mathcal{G}}+[\frac{1}{4}d_{1}d_{2}d_{3}
(\chi_{1}+\chi_{2})T^{d_{2}-1}\mathcal{G}^{\frac{1}{4}
(\chi_{1}+\chi_{2})-1}-\frac{1}{4}d_{2}^2d_{3}\right.\\\label{40}&&\left.\times(\chi_{1}-\chi_{2})
T^{d_{2}-1}\mathcal{G}^{-\frac{1}{4}(\chi_{1}-\chi_{2})-1}]\ddot{T})\right]
\geq0,\\\nonumber&&\textbf{DEC:}\quad\rho_{eff}-P_{eff}
=\rho+\frac{1}{\kappa^2}\left[[d_{1}d_{3}T^{d_{2}}\mathcal{G}^{\frac{1}{4}
(\chi_{1}+\chi_{2})}+d_{2}d_{3}T^{d_{2}}\right.\\\nonumber&&\left.\times
\mathcal{G}^{-\frac{1}{4}(\chi_{1}-\chi_{2})}-\chi_{3}T+d_{1}d_{2}
T^{\chi_{4}}+\chi_{5}T^{\chi_{6}}]+\rho[d_{1}d_{2}d_{3}T^{d_{2}-1}
\right.\\\nonumber&&\left.\times\mathcal{G}^{\frac{1}{4}(\chi_{1}+\chi_{2})}
+d_{2}^2d_{3}T^{d_{2}-1}\mathcal{G}^{-\frac{1}{4}(\chi_{1}-\chi_{2})}
-\chi_{3}+d_{1}d_{2}\chi_{4}T^{\chi_{4}-1}+\chi_{5}\chi_{6}T^{\chi_{6}-1}]
\right.\\\nonumber&&\left.+24qH^4[\frac{1}{4}d_{1}d_{3}(\chi_{1}+\chi_{2})
T^{d_{2}}\mathcal{G}^{\frac{1}{4}(\chi_{1}+\chi_{2})-1}
-\frac{1}{4}d_{2}d_{3}(\chi_{1}-\chi_{2})T^{d_{2}}\right.
\\\nonumber&&\left.\times\mathcal{G}^{-\frac{1}{4}(\chi_{1}-\chi_{2})-1}]+4H^3(3-2q)
([\frac{1}{4}d_{1}d_{3}(\chi_{1}+\chi_{2})\left[\frac{1}{4}(\chi_{1}
+\chi_{2})-1\right]\right.\\\nonumber&&\left.\times
T^{d_{2}}\mathcal{G}^{\frac{1}{4}(\chi_{1}+\chi_{2})-2}
+\frac{1}{4}d_{2}d_{3}(\chi_{1}-\chi_{2})
\left[\frac{1}{4}(\chi_{1}-\chi_{2})+1\right]T^{d_{2}}
\mathcal{G}^{-\frac{1}{4}(\chi_{1}-\chi_{2})-2}]\right.\\\nonumber&&\left.
\times\dot{\mathcal{G}}+[\frac{1}{4}d_{1}d_{2}d_{3}(\chi_{1}
+\chi_{2})T^{d_{2}-1}\mathcal{G}^{\frac{1}{4}(\chi_{1}+\chi_{2})-1}-\frac{1}{4}
d_{2}^2d_{3}(\chi_{1}-\chi_{2})T^{d_{2}-1}\right.\\\nonumber&&\left.\times
\mathcal{G}^{-\frac{1}{4}(\chi_{1}-\chi_{2})-1}]\dot{T})
+4H^2([\frac{1}{4}d_{1}d_{3}(\chi_{1}+\chi_{2})\left[\frac{1}{4}
(\chi_{1}+\chi_{2})-1\right]\right.\\\nonumber&&\left.\times
\left[\frac{1}{4}(\chi_{1}+\chi_{2})-2\right]T^{d_{2}}
\mathcal{G}^{\frac{1}{4}(\chi_{1}+\chi_{2})-3}-\frac{1}{4}d_{2}d_{3}
(\chi_{1}-\chi_{2})\left[\frac{1}{4}(\chi_{1}-\chi_{2})
\right.\right.\\\nonumber&&\left.\left.+1\right]\left[\frac{1}{4}
(\chi_{1}-\chi_{2})+2\right]T^{d_{2}}\mathcal{G}^{-\frac{1}{4}
(\chi_{1}-\chi_{2})-3}]\dot{\mathcal{G}}^2+2[\frac{1}{4}d_{1}d_{2}
d_{3}(\chi_{1}+\chi_{2})\right.\\\nonumber&&\left.\times\left[\frac{1}{4}
(\chi_{1}+\chi_{2})-1\right]T^{d_{2}-1}\mathcal{G}^{\frac{1}{4}(\chi_{1}
+\chi_{2})-2}+\frac{1}{4}d_{2}^2d_{3}(\chi_{1}-\chi_{2})
\left[\frac{1}{4}(\chi_{1}-\chi_{2})\right.\right.\\\nonumber&&\left.
\left.+1\right]T^{d_{2}-1}\mathcal{G}^{-\frac{1}{4}(\chi_{1}-\chi_{2})-2}]
\dot{\mathcal{G}}\dot{T}+[\frac{1}{4}d_{1}d_{2}d_{3}(d_{2}-1)
(\chi_{1}+\chi_{2})\right.\\\nonumber&&\left.\times
T^{d_{2}-2}\mathcal{G}^{\frac{1}{4}(\chi_{1}+\chi_{2})-1}
-\frac{1}{4}d_{2}^2d_{3}(d_{2}-1)(\chi_{1}-\chi_{2})T^{d_{2}-2}
\mathcal{G}^{-\frac{1}{4}(\chi_{1}-\chi_{2})-1}]\dot{T}^2
\right.\\\nonumber&&\left.+[\frac{1}{4}d_{1}d_{3}(\chi_{1}+\chi_{2})
\left[\frac{1}{4}(\chi_{1}+\chi_{2})-1\right]T^{d_{2}}\mathcal{G}^{\frac{1}{4}
(\chi_{1}+\chi_{2})-2}+\frac{1}{4}d_{2}d_{3}(\chi_{1}-\chi_{2})
\right.\\\nonumber&&\left.\times\left[\frac{1}{4}(\chi_{1}-\chi_{2})+1\right]
T^{d_{2}}\mathcal{G}^{-\frac{1}{4}(\chi_{1}-\chi_{2})-2}]
\ddot{\mathcal{G}}+[\frac{1}{4}d_{1}d_{2}d_{3}(\chi_{1}+\chi_{2})T^{d_{2}-1}
\right.\\\label{41}&&\left.\times\mathcal{G}^{\frac{1}{4}
(\chi_{1}+\chi_{2})-1}-\frac{1}{4}d_{2}^2d_{3}(\chi_{1}-\chi_{2})
T^{d_{2}-1}\mathcal{G}^{-\frac{1}{4}(\chi_{1}-\chi_{2})-1}]\ddot{T})\right]\geq0.
\end{eqnarray}
\begin{figure}
\epsfig{file=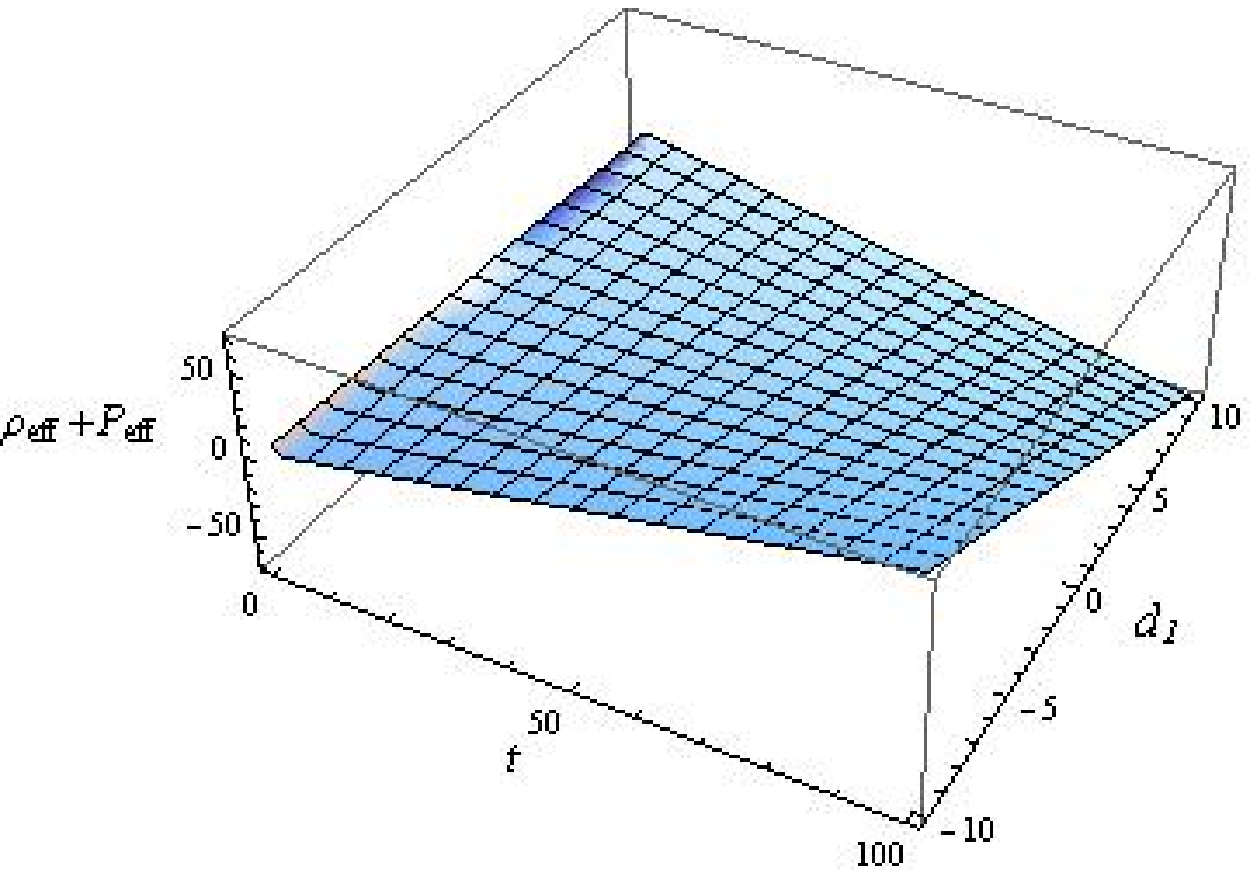, width=0.5\linewidth}\epsfig{file=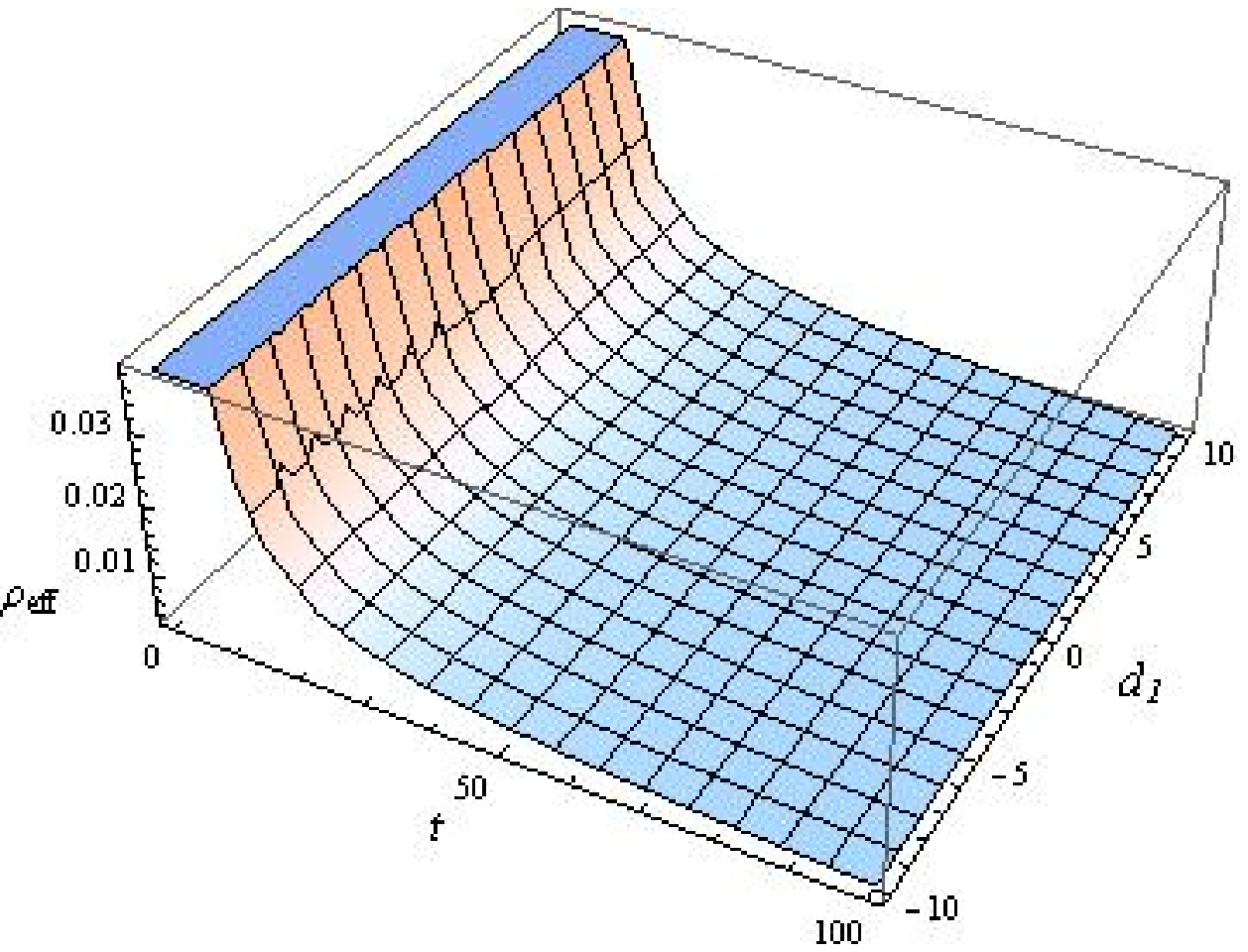,
width=0.5\linewidth}\caption{Energy conditions for $d_{2}=0.1$ and
$d_{3}=1$.}
\end{figure}
\begin{figure}
\epsfig{file=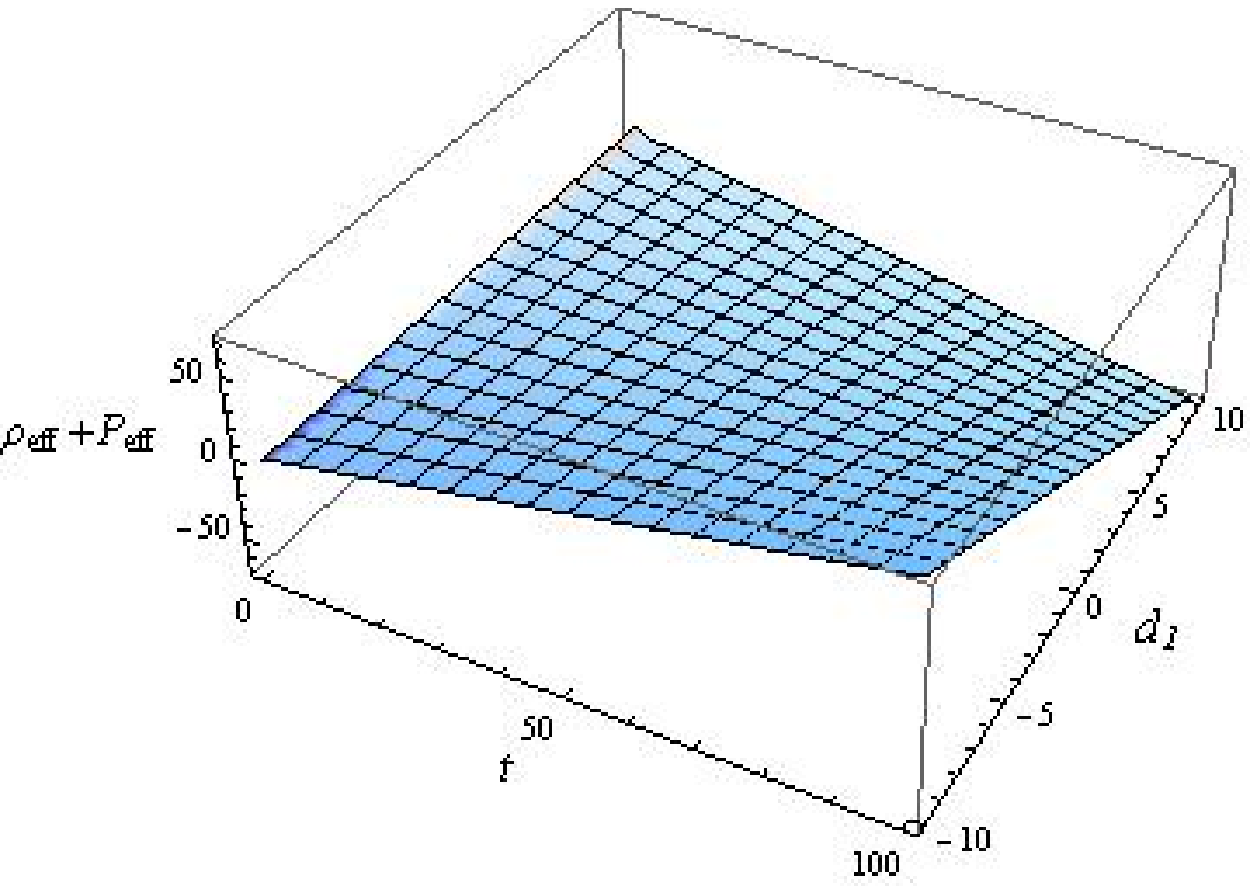, width=0.5\linewidth}\epsfig{file=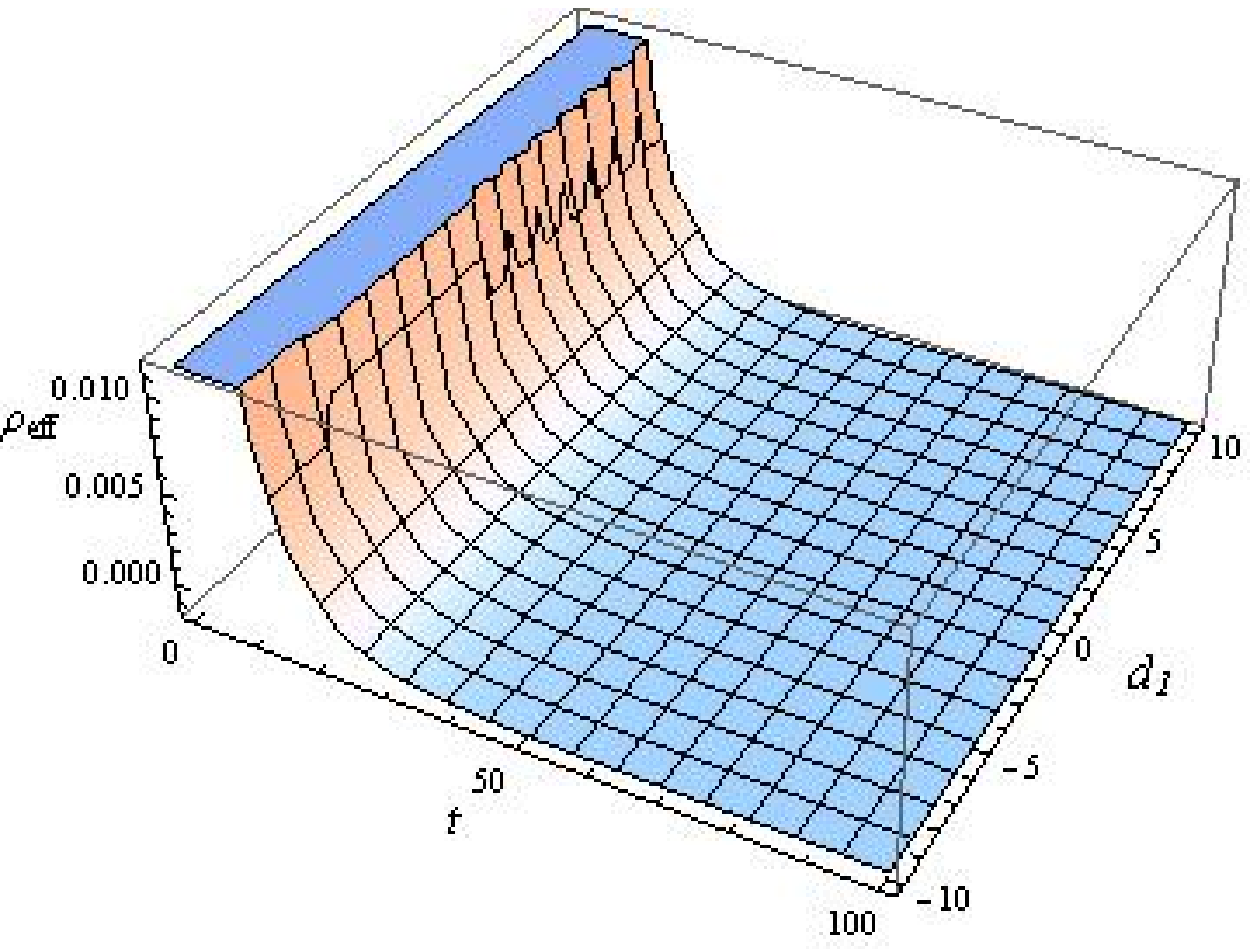,
width=0.5\linewidth}\caption{Energy conditions for $d_{2}=0.1$ and
$d_{3}=-0.5$.}
\end{figure}
\begin{figure}
\epsfig{file=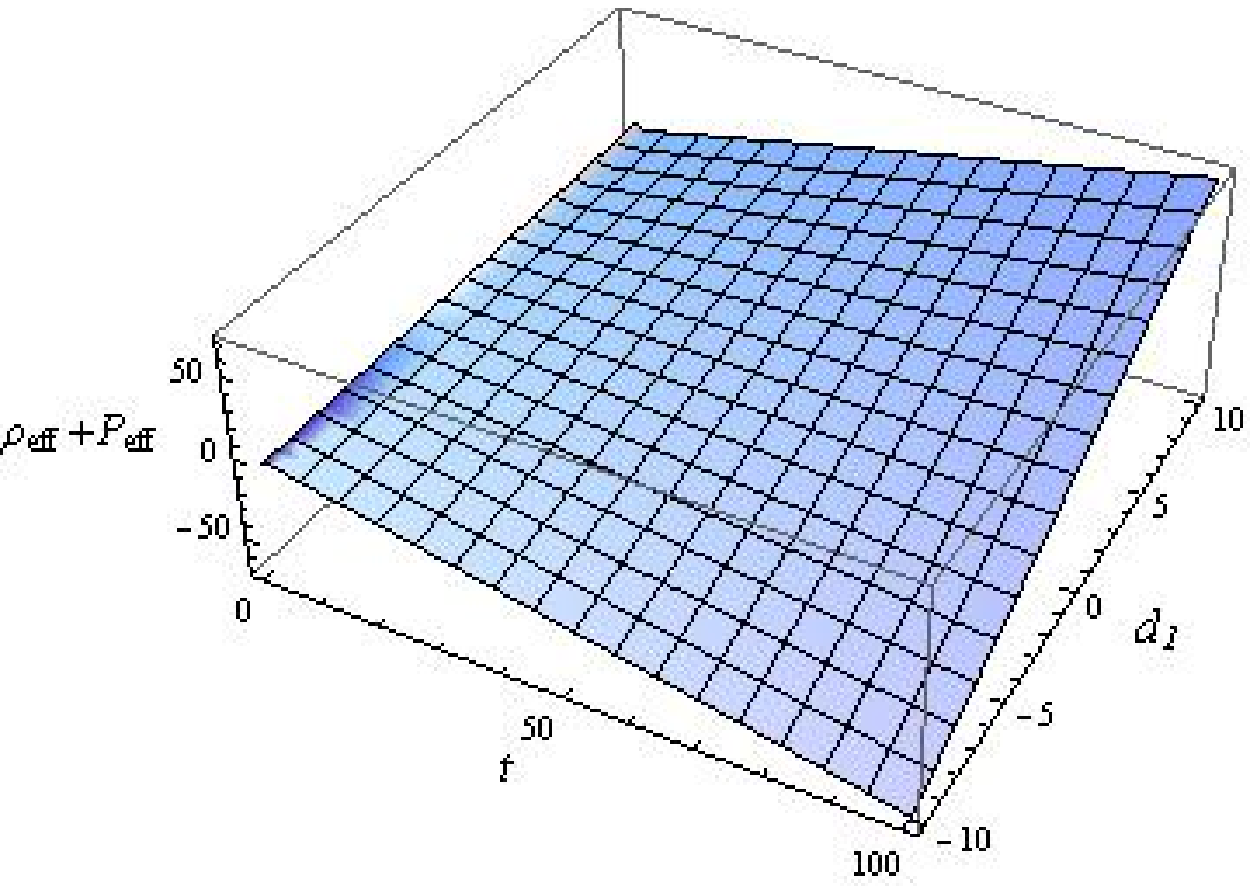, width=0.5\linewidth}\epsfig{file=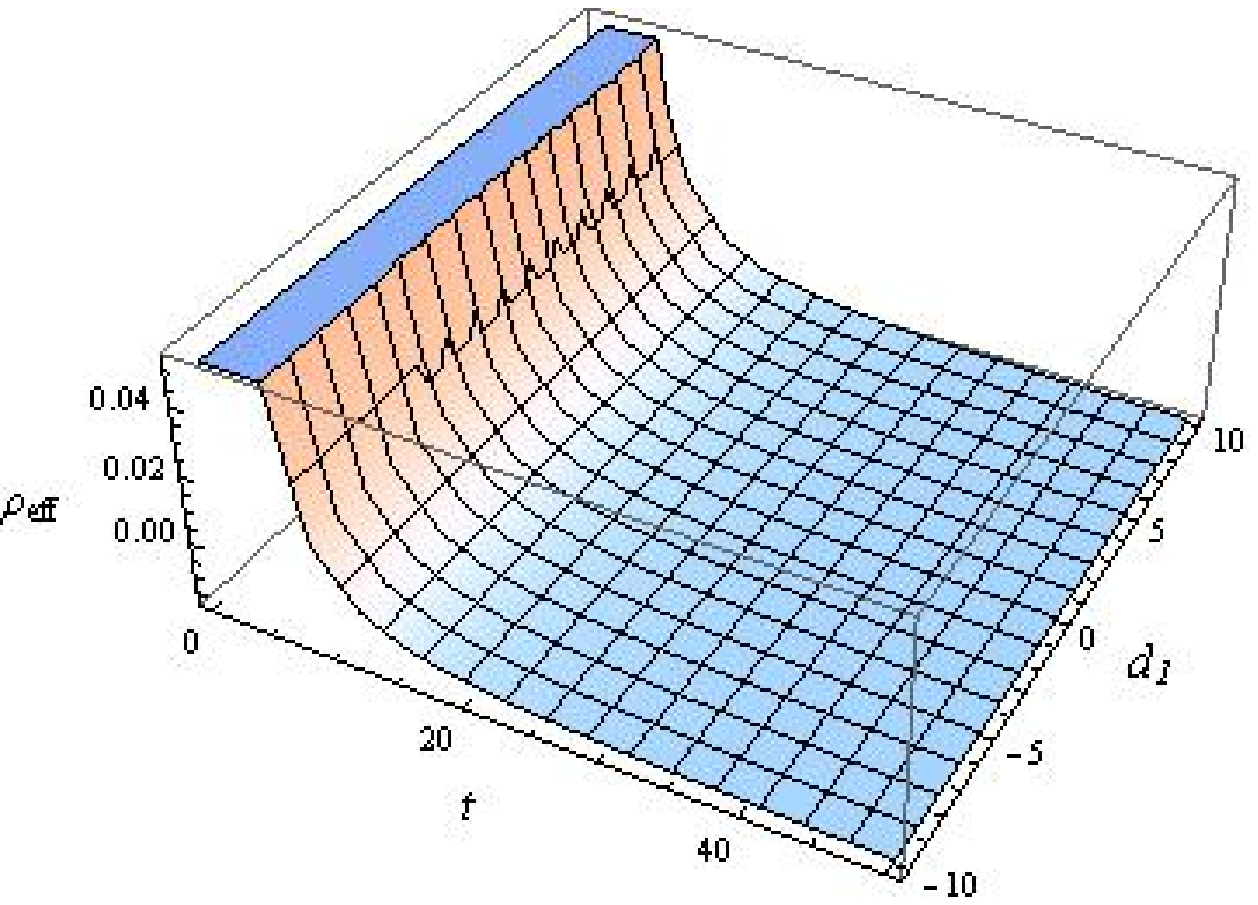,
width=0.5\linewidth}\caption{Energy conditions for $d_{2}=-0.1$ and
$d_{3}=0.5$.}
\end{figure}
\begin{figure}
\epsfig{file=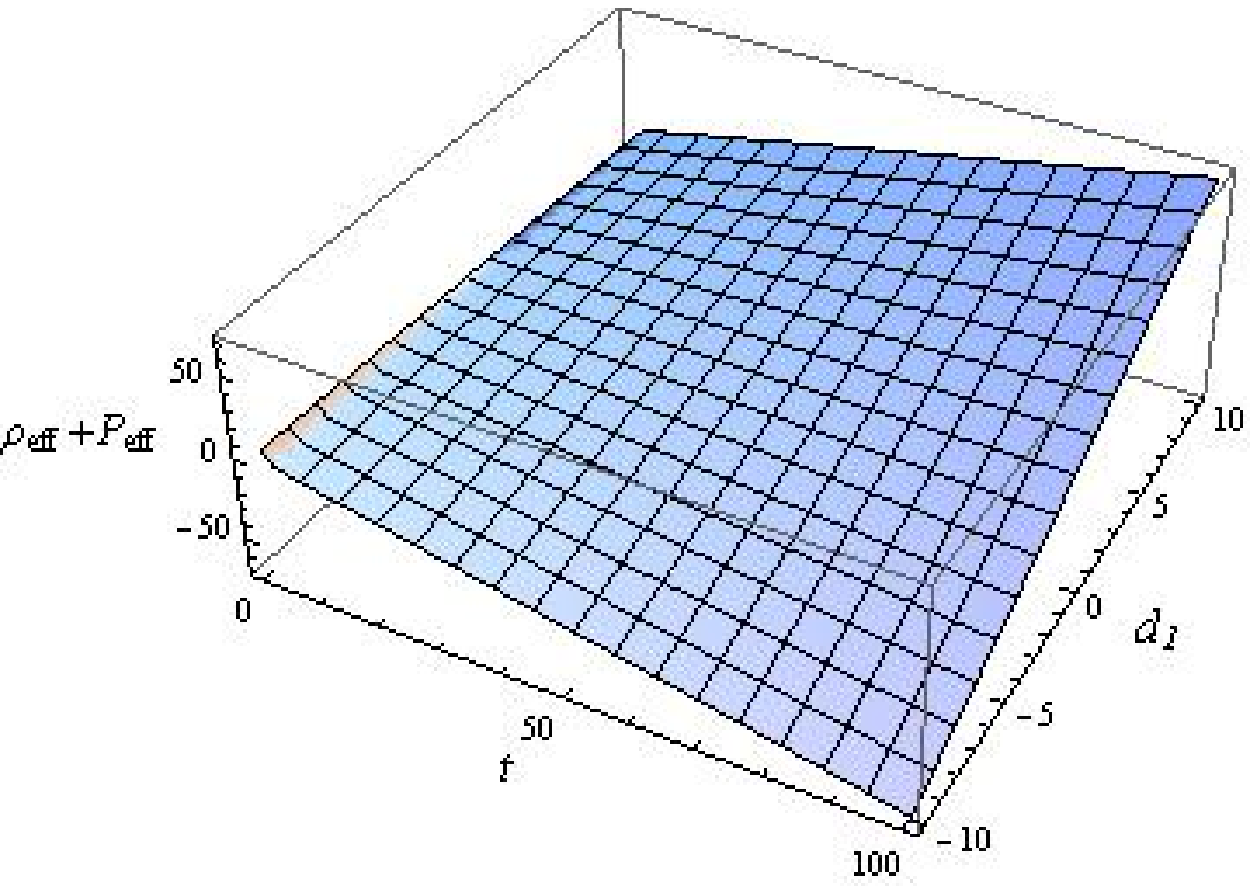, width=0.5\linewidth}\epsfig{file=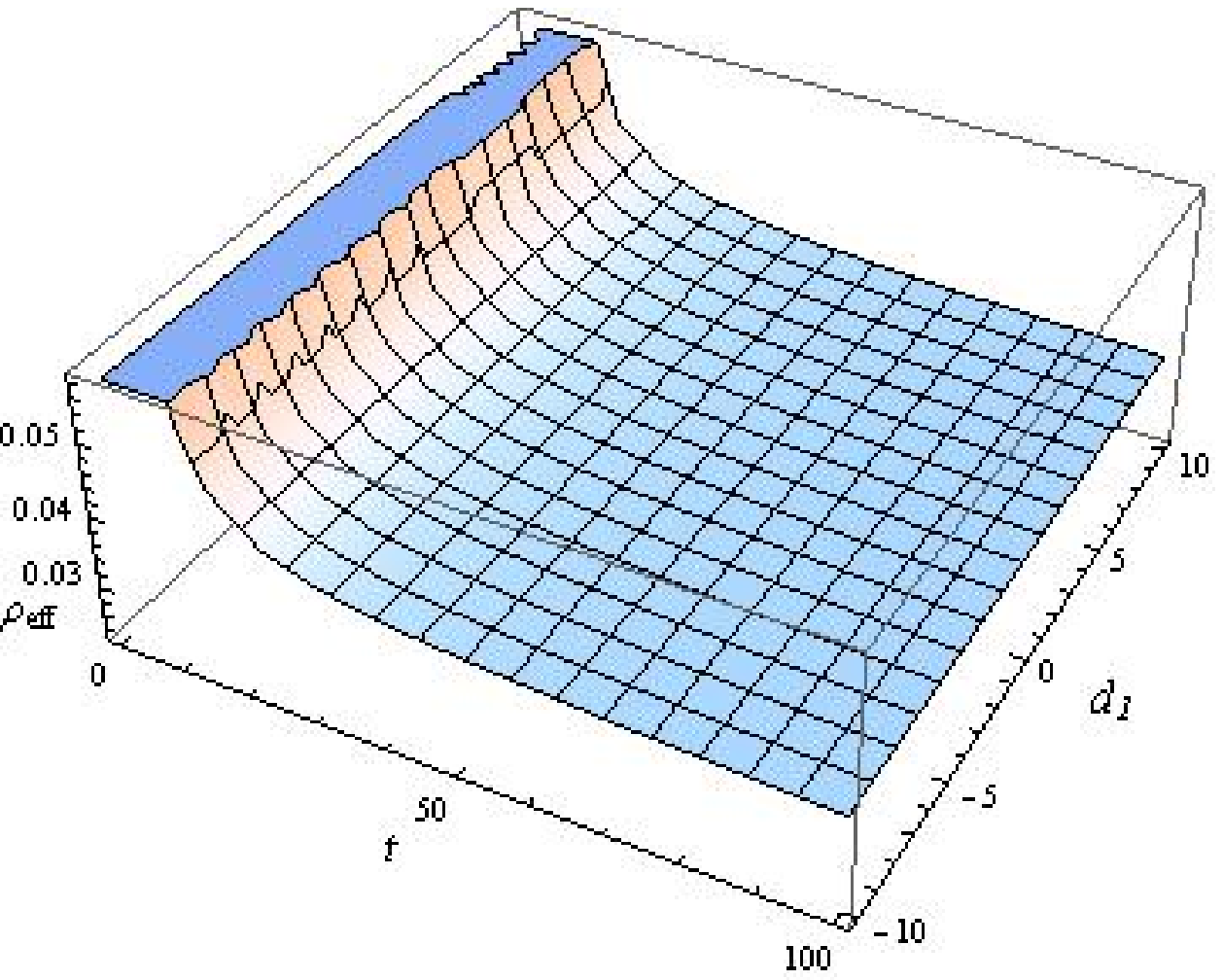,
width=0.5\linewidth}\caption{Energy conditions for $d_{2}=-0.1$ and
$d_{3}=-1$.}
\end{figure}

The NEC and WEC depend on four parameters $t,~d_{1},~d_{2}$ and
$d_{3}$. We plot these conditions against $t$ and $d_{1}$ for
$n=\frac{2}{3}$ with possible signs of $d_{2}$ and $d_{3}$. The left
plot of Figure \textbf{9} shows positively increasing behavior of
NEC for $-10\leq d_{1}\leq0$ with respect to time while invalid for
$d_{1}>0$. The effective energy density remains positive for all
values of $(t,d_{1})$ as shown in Figure \textbf{9} (right). The
same behavior of both conditions are obtained for $0<d_{2}\leq0.51$
with $d_{3}>0$ as well as for $d_{2}>0$ with $d_{3}=0$. The left
plot of Figure \textbf{10} shows similar behavior of NEC for
$d_{2}>0$ and $d_{3}<0$ while $\rho_{eff}$ remains positive for
$0<t<23$. Similarly, for $d_{3}=-1$ and $-10$, WEC is valid for
$0<t<14$ and $0<t<4.5$, respectively with $d_{2}=0.1$. The right
plot of Figures \textbf{11} and \textbf{12} shows the validity of
NEC for $d_{1}\geq0$ while does not hold for negative values of
$d_{1}$. The effective energy density remains positive for time
interval $1\leq t\leq 10$ with $d_{3}=0.5$ as shown in Figure
\textbf{11} (right panel) while for $d_{3}=1$ and $10$, the
acceptable intervals are $1\leq t\leq 7$ and $1\leq t\leq 3$,
respectively. This shows that the validity region of WEC decreases
as the value of integration constant $d_{3}$ increases. The right
plot of Figure \textbf{12} shows the positivity of $\rho_{eff}$ for
$(d_{2},d_{3})<0$ which confirms the positivity of WEC with
$d_{1}>0$.

\section{Final Remarks}

In this paper, we have presented a generalized modified theory of
gravity with an arbitrary coupling between geometry and matter. The
gravitational Lagrangian is obtained by adding an arbitrary function
$f(\mathcal{G},T)$ in the Einstein-Hilbert action. We have
formulated the corresponding field equations using least action
principle and calculated the non-zero covariant divergence of
$T_{\alpha\beta}$ consistent with $f(R,T)$ theory \cite{17}.
Consequently, the test particles follow non-geodesic trajectories
due to the presence of extra force originated from the non-minimally
coupling while they move along geodesics for pressureless fluid. We
have constructed energy conditions for FRW universe model filled
with dust fluid in terms of deceleration, jerk and snap $(q,j,s)$
cosmological parameters. The reconstruction technique has been
applied to $f(\mathcal{G},T)$ gravity using well-known de Sitter and
power-law universe models. The results are summarized as follows.
\begin{itemize}
\item In de Sitter reconstructed model, the energy bounds have dependence
on three parameters $t,~c_{1}$ and $c_{2}$. We have plotted NEC and
WEC against $t$ and $c_{2}$ with four possible signatures of $c_{1}$
and $c_{2}$ as shown in Figures \textbf{1}-\textbf{8}. It is found
that NEC and WEC are satisfied for $c_{1}>0$ and $c_{2}<0$
throughout the time interval while for cases $(c_{1},c_{2})>0$ and
$(c_{1},c_{2})<0$, energy conditions are satisfied for small values
of $c_{i}$'s in a very small time interval. It is observed that the
NEC shows positively increasing behavior for all negative values of
$c_{1}$ with $c_{2}>0$ while the validity ranges of WEC have
dependence on $c_{1}$.
\item For power-law reconstructed model, we have explored the behavior of
four parameters $t,~d_{1},~d_{2}$ and $d_{3}$ with $n=\frac{2}{3}$.
In this case, we have plotted energy conditions against $(t,d_{1})$
and analyzed possible behavior of remaining constants. In Figures
\textbf{9}-\textbf{12}, we have taken $-10\leq d_{1}\leq 10$ and
found the valid regions where energy conditions are satisfied.
\end{itemize}
Finally, we conclude that the NEC and WEC are satisfied in both
reconstructed $f(\mathcal{G},T)$ models with suitable choice of free
parameters.

\end{document}